\documentclass[twocolumn,showpacs,preprintnumbers,
pra,amsmath,amssymb,superscriptaddress]{revtex4}

\usepackage{bm}
\usepackage{epsfig}

\newcommand{\bfsfG}{\mbox{\sffamily\bfseries{G}}}
\newcommand{\bfsfI}{\mbox{\sffamily\bfseries{I}}}
\newcommand{\bfsfK}{\mbox{\sffamily\bfseries{K}}}

\newcommand{\bfsfT}{\mbox{\sffamily\bfseries{T}}}
\newcommand{\bfsfV}{\mbox{\sffamily\bfseries{V}}}
\newcommand{\bfmu}{\bm \mu}

\begin{document}
\title{A multiple-scattering approach to interatomic interactions and superradiance in inhomogeneous
dielectrics}
\date{To appear in Physical Review A}
\author{Martijn Wubs} \email{Martijn.Wubs@physik.uni-augsburg.de}
\homepage{http://tnweb.tn.utwente.nl/cops/} \affiliation{Complex
Photonic Systems, Faculty of Science and Technology, University of
Twente, P.O. Box 217, NL-7500~AE~~Enschede, The Netherlands}
\affiliation{Van der Waals-Zeeman Institute, University of
Amsterdam, Valckenierstraat 65, NL-1018 XE  Amsterdam, The
Netherlands}
\author{L.G.~Suttorp}
\affiliation{Institute for Theoretical Physics, University of
Amsterdam, Valckenierstraat 65, NL-1018 XE  Amsterdam, The
Netherlands}
\author{A.~Lagendijk}
\affiliation{Complex Photonic Systems, Faculty of Science and
Technology, University of Twente, P.O. Box 217,
NL-7500~AE~~Enschede, The Netherlands}

\begin{abstract}
The dynamics of a collection of resonant atoms embedded inside an
inhomogeneous nondispersive and lossless dielectric is described
with a dipole Hamiltonian that is based on a canonical
quantization theory. The dielectric is described macroscopically
by a position-dependent dielectric function and the atoms as
microscopic harmonic oscillators. We identify and discuss the role
of several types of Green tensors that describe the
spatio-temporal propagation of field operators. After integrating
out the atomic degrees of freedom, a multiple-scattering formalism
emerges in which an exact Lippmann-Schwinger equation for the
electric field operator plays a central role. The equation
describes atoms as point sources and point scatterers for light.
First, single-atom properties are calculated such as
position-dependent spontaneous-emission rates as well as
differential cross sections for elastic scattering and for
resonance fluorescence. Secondly, multi-atom processes are
studied. It is shown that the medium modifies both the resonant
and the static parts of the dipole-dipole interactions.  These
interatomic interactions may cause the atoms to scatter and emit
light cooperatively. Unlike in free space,  differences in
position-dependent emission rates and radiative line shifts
influence cooperative decay in the dielectric. As a generic
example, it is shown that near a partially reflecting plane there
is a sharp transition from two-atom superradiance to single-atom
emission as the atomic positions are varied.
\end{abstract}


\maketitle

\section{Introduction}\label{chapquantcorrintro}
The spontaneous-emission rate of an atom depends on its dielectric
environment \cite{Purcell46,Nienhuis76} and in particular on the
precise position of the atom if the medium is inhomogeneous
\cite{Drexhage70,Agarwal75,Glauber91,Khosravi91,Suzuki95,Snoeks95}.
Spontaneous emission can only be understood quantum mechanically,
but the classical Green function determines the emission rate. In
particular, the emission rate is proportional to the imaginary
part of the Green tensor of the medium at the atomic position. The
dipole-angle average of the emission rate is also known as the
local optical density of states \cite{Sprik96}.

In order to study the influence of the medium on much more than
just single-atom spontaneous-emission rates, in this paper a
rather general multiple-scattering theory is set up. It is is
based on `macroscopic quantization' theories of the
electromagnetic field in inhomogeneous lossless dielectrics, see
for example \cite{Knoell87,Glauber91,Dalton,Wubs03}. Microscopic
treatments of (macroscopically homogeneous) dielectrics  in
quantum electrodynamics can be found in
 \cite{Knoester89,Juzeliunas94} but will not be used here.
 An important result in the macroscopic  theories is that
photons can be defined as the elementary excitations of the true
modes of the dielectric. Corresponding mode functions are the
(classical) harmonic solutions of the wave equation.

Emission rates of an atom not only change due to the nonresonant
dielectric environment, but also due to the presence of other
atoms with the same transition frequency. As is known since the
pioneering work by Dicke \cite{Dicke54}, resonant atoms in each
others neighborhood decay cooperatively. Depending on the
many-atom state, the atoms decay faster than a single atom up till
twice the single-atom rate  (superradiance) or decay slower or not
at all (subradiance). Lifetime changes of individual atom pairs as
a function of their distance were measured only recently
\cite{DeVoe96}; for two ${\rm Ba}^{+}$ ions that emit at  a
wavelength of $493\;{\rm nm}$ and for well-defined separations
$|{\bf R}|$ around $1.5\;\mu{\rm m}$, subradiant and superradiant
lifetime effects of less than $\pm 2\%$ were observed.

Superradiance occurs for the so-called Dicke states that have a
zero expectation value of the total dipole moment \cite{Dicke54},
but also for atomic product states with a nonzero dipole moment
\cite{Allen75,Mandel95}. Superradiance also occurs for classical
dipoles. It is a general phenomenon also exhibited in acoustics by
nearby identical tuning forks, or by strings in a piano
\cite{Mandel95}. These systems have in common that the atoms (or
oscillators) interact with a field that is influenced by the
radiation reactions of all nearby atoms together.

Cooperative effects of resonant atoms will be influenced by their
dielectric environment.  In this paper the influence of a
nondispersive and lossless inhomogeneous dielectric on embedded or
nearby resonant atoms is studied. Quantum theory is used both to
describe the light and the atoms. As we focus on the effects of
the inhomogeneous dielectric, the atoms are modelled simply as
quantum harmonic oscillators in their ground states or first
excited states, with fixed dipole orientations. To be sure, in
choosing this model we neglect optical saturation effects of the
atoms. The dielectric is described macroscopically in terms of a
real-valued relative dielectric function $\varepsilon({\bf r})$,
the form of which will be left arbitrary. The precise measurement
of two-atom superradiance in free space as a function of distance
is a fundamental test for quantum electrodynamics \cite{DeVoe96}.
The effects calculated here are a test for macroscopic
quantization theories for inhomogeneous dielectrics.

Like single-atom emission \cite{DeMartini91}, two-atom
superradiance will be modified in the close vicinity of a mirror,
or inside an optical cavity \cite{Ujihara02}. Strong modifications
of superradiance are also predicted for photonic crystals,
dielectrics with periodic refractive-index variations on the scale
of the wavelength of light \cite{Kurizki88,John95,Vats98}. Large
effects are predicted in the so-called isotropic model for a
photonic crystal, which is really a toy model in the sense that
all local an orientational inhomogeneities of the electromagnetic
field are neglected. In a real photonic crystal,
 two-atom superradiance  is expected to sensitively depend on the coordinates of
both atoms. The present  formalism is valid for an arbitrary real
dielectric function and encompasses the interesting special cases
just mentioned.

Atoms that exhibit superradiance interact strongly enough to share
and exchange the optical excitation before emission. The more
common and better studied situation for resonant atoms in a
dielectric is that the interaction between the atoms is weak
compared to interactions with baths that the individual atoms
have. Then the optical excitations are transferred irreversibly
from donor to acceptor atoms via a process called `resonance
energy transfer', as described by F{\" o}rster theory
\cite{Foerster48} and its modern generalizations \cite{Scholes03}.
Resonance energy transfer is influenced by the dielectric
environment. For example, calculations show that two-atom
interactions can be strongly influenced by an optical microcavity
\cite{Kobayashi95,Agarwal98,Hartman01}, since the cavity modes
with eigenfrequencies close or equal to the atomic transition
frequency play a dominant role. Indeed, experiments have shown
that the interatomic (dipole-dipole) interaction is increased when
the atoms are placed in a cavity at positions where resonant
optical modes
 have their maxima \cite{Hopmeier99}. In another interesting experiment,
  F{\"o}rster excitation transfer is found to scale linearly with
the local optical density
 of states at the donor position \cite{Andrew00}.
Although we focus on superradiance, the formalism in the present
work is quite general and can also be used as a quantum
electrodynamical foundation for the study of energy transfer
processes in inhomogeneous media. Recent progress in this
direction can be found in \cite{Dung02,Colas03}.

Often in quantum optics an `all-matter' picture is employed, where
the dynamics of the electromagnetic field is integrated out, for
example in the optical Bloch equations \cite{Allen75,Loudon83}.
Here instead we treat spontaneous emission and superradiance in an
`all light' picture, which is convenient when studying the effect
of the dielectric. A multiple-scattering theory is set up in which
the atoms show up both as sources and as scatterers of light.   It
is known that superradiance can be viewed as caused by
multiple-scattering interactions
\cite{Milonni74,Brewer95,Lawande90}. Light scattered off a
collection of atoms will show multi-atom resonances and
cooperative effects, also due to multiple scattering. In quantum
scattering theory such resonances  appear as well and sometimes
are called `proximity resonances' \cite{Heller96,Li03}.

  The
concept of a point scatterer proved very fruitful in the study of
multiple-scattering of classical light in free space
\cite{Lagendijk96,DeVries98a,DeVries98b}. Here, the
point-scattering formalism will be put to use in quantum optics of
inhomogeneous dielectrics. Multiple light scattering will be
described in terms of Green functions of the medium. The emphasis
of the paper will be on formalism, but it ends with an application
 to superradiance in a  model dielectric.

 The paper has the following structure: in
Sec.~\ref{scatsesingle} the point-scattering model for interacting
guest atoms is introduced. Properties of several types of Green
functions of the medium are discussed in Sec.~\ref{Greenmedium}.
Sec.~\ref{Singlepoint} discusses medium-induced modifications of
single-atom properties such as spontaneous-emission rates and
elastic scattering. The generalization to a finite number of host
atoms is discussed in Sec.~\ref{severalsources}. The formalism is
applied to two-atom superradiance in Sec.~\ref{supertwo}, in
particular to superradiance near a partially reflecting plane in
Sec.~\ref{secsuperplane}. Conclusions are drawn in
Sec.~\ref{quantcorrconc}.

\section{Atoms as point sources and as point scatterers}
\label{scatsesingle}

\subsection{The Hamiltonian}\label{Hamiltonianseveralatoms}
Consider an inhomogeneous dielectric with relative dielectric
function $\varepsilon({\bf r})$ with a finite number $N$ of
embedded neutral atoms. The dipole Hamiltonian for this system is
the sum of a field part, an atomic part, and an interaction part
between field and atoms.   More precisely, the Hamiltonian can be
found after canonical quantization \cite{Dalton,Wubs03} to have
the form $\mathcal{H} = \mathcal{H}_{\rm F} + \mathcal{H}_{\rm A}
+ \mathcal{H}_{\rm AF}$, with
\begin{subequations}\label{Hafquantcorrglobal}
\begin{eqnarray}
\mathcal{H}_{\rm F} & = & \sum_\lambda \hbar \omega_{\lambda}
a_{\lambda}^{\dag}a_{\lambda}
\\ \mathcal{H}_{\rm A} & = &  \sum_{m=1}^{N}\hbar \Omega_{m} b_{m}^{\dag} b_{m} \\
\mathcal{H}_{\rm AF}& = &  -\sum_{m=1}^{N}{\bm \mu}_{m}\cdot {\bf
F}({\bf R}_{m}) \nonumber \\ & = & \sum_{m,\lambda}
(b_{m}+b_{m}^{\dag})(g_{\lambda m}a_{\lambda} + g_{\lambda
m}^{*}a_{\lambda}^{\dag}).\label{Hafquantcorr}
\end{eqnarray}
\end{subequations}
 Notice that
there is no direct interaction term between neutral atoms. In a
minimal-coupling Hamiltonian, there would have been such a direct
coupling term. The situation is analogous to the free-space case
\cite{Cohen89,Loudon83}. The field part $\mathcal{H}_{\rm F}$ of
the Hamiltonian is a sum (or integral) over harmonic oscillators
corresponding to the harmonic solutions (`true modes') ${\bf
f}_\lambda$ of the Maxwell equations for the inhomogeneous
dielectric in the absence of the atoms:
\begin{equation}\label{true modes}
-\nabla\times\nabla\times {\bf f}_{\lambda}({\bf r})
+\varepsilon({\bf r})(\omega_{\lambda}/c)^{2}{\bf
f}_{\lambda}({\bf r}) = 0.
\end{equation}
For $\omega_{\lambda} \ne 0$, these modes are generalized
transverse, which means that $\nabla \cdot [\varepsilon({\bf r})
{\bf f}_{\lambda}({\bf r}) ]\equiv 0$. Their orthonormality
condition reads $\int\mbox{d}{\bf r}\varepsilon({\bf r}){\bf
f}_{\lambda}^{*}({\bf r})\cdot{\bf f}_{\lambda'}({\bf r}) =
\delta_{\lambda\lambda'}$, where $*$ denotes complex conjugation.
The modes are complete, in other words they form a basis for the
subspace of generalized transverse functions. For free space
[$\varepsilon({\bf r})\equiv 1$] the ${\bf f}_{\lambda}$ are the
well-known transverse plane-wave modes.

In the atomic Hamiltonian $\mathcal{H}_{\rm A}$, the atomic
transition frequencies $\Omega_{m}$ and  transition dipole moments
${\bm \mu}_{m}$ may be all different, either because the guest
atoms are of  different species or because identical atoms feel a
different environment. The frequencies $\Omega_{m}$ are assumed
real, which means that nonradiative broadening is neglected. The
atoms are very simply described as harmonic oscillators with
frequencies $\Omega_{m}$. This is a good approximation within a
certain frequency range and as long as saturation effects of the
upper atomic state can be neglected. The atomic transition dipole
moments ${\bm \mu}_{m}$ are assumed to be real-valued and to have
fixed orientations. This assumption is better for
 molecules or quantum dots in a solid surrounding than for atoms in the gas
phase. For convenience, the name `atoms' will be used for the
guests in the dielectric.  The operators $b_{m}^{\dag}(t)$ create
atomic excitations by annihilating an atom in the ground state
while at the same time creating the atom in the excited state.

The total displacement field ${\bf D}({\bf r},t)$ is equal to the
displacement field $\varepsilon_{0}\varepsilon({\bf r}){\bf
E}({\bf r},t)$ of the inhomogeneous medium plus the sum
$\sum_{m}{\bf P}_{m}({\bf r},t)$ of the polarization fields
produced by the guest atoms. In the dipole approximation, these
polarization fields have the form
\begin{eqnarray}\label{pgpoint}
{\bf P}_{m}({\bf r},t) & = & \delta({\bf r}-{\bf R}_{m})\;{\bf
P}_{m}(t) \nonumber \\ &  = & \delta({\bf r}-{\bf R}_{m})\;{\bm
\mu}_{m}\left[\;b_{m}(t)+b_{m}^{\dag}(t)\;\right].
\end{eqnarray}
In the dipole interaction term $\mathcal{H}_{\rm AF}$ of the
Hamiltonian, a field called ${\bf F}$ was introduced that is an
abbreviation of
\begin{equation}\label{Fdef}
{\bf F}({\bf r},t)\equiv {\bf D}({\bf
r},t)/[\varepsilon_{0}\varepsilon({\bf r})].
\end{equation}
Atomic dipoles couple to this field ${\bf F}({\bf r},t)$
\cite{Dalton,Wubs03}. It is equal to the electric field operator
${\bf E}({\bf r},t)$ everywhere, except at the positions ${\bf
R}_{m}$ of the guests, since the guest dipoles couple  to fields
in which their own polarization fields are included. For free
space this self-interaction in the dipole coupling is known
\cite{Cohen89}. The mode expansion of the field ${\bf F}({\bf
r},t)$ has a simple form, being the sum of a positive-frequency
part ${\bf F}^{(+)}({\bf r},t)$ containing only annihilation
operators and its Hermitian conjugate ${\bf F}^{(-)}({\bf r},t)$,
where
\begin{equation}\label{ddooreps}
{\bf F}^{(+)}({\bf r},t)= i \sum_{\lambda} \sqrt{\frac{\hbar
\omega_{\lambda}}{2 \varepsilon_{0}}} a_{\lambda}(t)\; {\bf
f}_\lambda({\bf r}).
\end{equation}
In the absence of the atoms, the time dependence of the
annihilation operators in (\ref{ddooreps}) would be harmonic and
${\bf F}({\bf r},t)$ would be equal to the electric field ${\bf
E}^{(0)}({\bf r},t)$. Here and below, the superscript $(0)$
denotes the absence of guest atoms in the inhomogeneous
dielectric. For convenience, coupling constants between atom $m$
and optical mode $\lambda$ in Eq.~(\ref{Hafquantcorr}) are defined
as
\begin{equation}\label{gkj}
g_{\lambda m} = - i \sqrt{\frac{\hbar \omega_{\lambda} }{2
\varepsilon_{0} } } {\bm \mu}_{m}\cdot {\bf f}_{\lambda}({\bf
R}_{m}).
\end{equation}
Notice that the coupling constants $g_{\lambda m}$ are zero for
(longitudinal) modes corresponding to $\omega_{\lambda}=0$. It is
by a convenient choice of gauge that the longitudinal modes are
decoupled from the atoms in the
Hamiltonian~(\ref{Hafquantcorrglobal}).

\subsection{Derivation of Lippmann-Schwinger equation}\label{intoutatomdyn}
The goal of this section is to derive a Lippmann-Schwinger
equation for the field ${\bf F}$ inside the inhomogeneous
dielectric in the presence of the $N$ guest atoms, by integrating
out the atomic dynamics. Heisenberg's equation of motion leads to
the following equations of motion for the field operators:
\begin{subequations}\label{aandbminsecondorder}
\begin{eqnarray}
 \dot{a}_\lambda &  = &  -i\omega_{\lambda}
a_\lambda - (i/\hbar)\sum_{m}g_{\lambda m}^{*}( b_{m} +
b_{m}^{\dag} ) \label{aformoredotdot}\\
 \dot{a}_{\lambda}^{\dag} &  = &  i\omega_{\lambda}
a_{\lambda}^{\dag} + (i/\hbar)\sum_{m}g_{\lambda m}( b_{m} +
b_{m}^{\dag}). \label{adagformoredotdot}
\end{eqnarray}
\end{subequations}
(The dot denotes the time derivative; explicit time dependence of
the operators is henceforth dropped.) The field operators are
coupled to the atomic operators and the operators of atom $m$
satisfy the equations
\begin{subequations}\label{atomicvars}
\begin{eqnarray} \dot{b}_{m}&  =
&-i\Omega_{m}b_{m} - (i/\hbar) \sum_{\lambda}( g_{\lambda
m}a_{\lambda} + g_{\lambda
m}^{*} a_{\lambda m}^{\dag} ) \label{dotb} \\
\dot{b}_{m}^{\dag} &  = & i\Omega_{m}b_{m}^{\dag} + (i/\hbar)
\sum_{\lambda} ( g_{\lambda m}a_{\lambda} + g_{\lambda m}^{*}
a_{\lambda m}^{\dag} ). \label{dotbdag}
\end{eqnarray}
\end{subequations}
Now take the Laplace transform (or one-sided Fourier transform) of
the equations of motion. The transform  will have the  argument
$-i\omega$, for example $b_{m}(\omega) \equiv
\int_{0}^{\infty}\mbox{d}t\;e^{i \omega t} b_{m}(t)$. Here and in
the following the frequency $\omega$ is assumed to contain an
infinitesimally small positive imaginary part so that the
transform is well-defined. The equations are algebraic after the
transformation.

 Also in  Fourier language, the equations
for the frequency-dependent atomic operators become
\begin{subequations}\label{dotbandbdagfourier}
\begin{eqnarray}
b_{m}(\omega)&  = &\frac{i b_{m}(t=0)}{\omega - \Omega_{m}}
\nonumber \\& +& \frac{\hbar^{-1}}{\omega-\Omega_{m}}
\sum_{\lambda} \left[ g_{\lambda m}a_{\lambda}(\omega) +
g_{\lambda
m}^{*} a^{\dag}_{\lambda m}(\omega) \right] \label{dotbfourier} \\
b^{\dag}_{m}(\omega) &  = &\frac{i b_{m}^{\dag}(t=0)}{\omega +
\Omega_{m}} \nonumber \\ &-& \frac{\hbar^{-1}}{\omega+\Omega_{m}}
\sum_{\lambda} \left[ g_{\lambda m}a_{\lambda}(\omega) +
g_{\lambda m}^{*} a^{\dag}_{\lambda m}(\omega)\right]
\label{dotbdagfourier}.
\end{eqnarray}
\end{subequations}
  In obtaining these equations, it was assumed that at
time zero, the annihilation operators $a_{\lambda}(t)$ coincide
with the $a_{\lambda}^{(0)}(t)$, the operators in the absence of
the guest atoms. The latter operators have the simple harmonic
time dependence
$\dot{a}_{\lambda}^{(0)}(t)+i\omega_{\lambda}a_{\lambda}^{(0)}(t)=0$,
 the    transform of which becomes
$-i(\omega-\omega_{\lambda})\;a_{\lambda}^{(0)}(\omega) =
a_{\lambda}^{(0)}(t=0)$ after a partial integration. Notice that
$b_{m}(\omega)$ and $b^{\dag}_{m}(\omega)$ in
Eq.~(\ref{dotbandbdagfourier}) are defined as the transforms of
$b_{m}(t)$ and $b_{m}^{\dag}(t)$, respectively. The time-dependent
operators are
 hermitian conjugates ($b_{m}^{\dag}(t)\equiv \left[b_{m}(t)\right]^\dag)$, but the frequency-dependent operators are not
($b^{\dag}_{m}(\omega)\ne \left[b_{m}(\omega)\right]^{\dag}$).

The right-hand sides of Eqs.~(\ref{dotbfourier}) and
(\ref{dotbdagfourier}) will now be used to replace $b_{m}(\omega)$
and $b^{\dag}_{m}(\omega)$ in the Laplace transforms of the
 Eqs. (\ref{aformoredotdot}) and (\ref{adagformoredotdot}) for
the field operators. In doing this, the atomic dynamics is
integrated out. One obtains for the frequency-dependent
annihilation and creation operators of the electromagnetic field
\begin{widetext}
\begin{subequations}\label{fieldoperatorsafterintegratingout}
\begin{eqnarray}\label{fieldannafterintegratingout}
a_{\lambda}(\omega) & = & a_{\lambda}^{(0)}(\omega) + \frac{i
\hbar^{-1}}{\omega-\omega_{\lambda}}\sum_{m} g_{\lambda
m}^{*}\left[\frac{b_{m}(0)}{\omega-\Omega_{m}}+
\frac{b_{m}^{\dag}(0)}{\omega+\Omega_{m}}\right]    +
\frac{\hbar^{-2}}{\omega-\omega_{\lambda}}
\sum_{m,\lambda'}\frac{2 g_{\lambda
m}^{*}\Omega_{m}}{\omega^{2}-\Omega_{m}^{2}} \left[ g_{\lambda'
m}\;a_{\lambda'}(\omega) + g_{\lambda'
m}^{*}\;a^{\dag}_{\lambda'}(\omega) \right], \\
a^{\dag}_{\lambda}(\omega) & = & a_{\lambda}^{(0)\dag}(\omega) -
\frac{i \hbar^{-1}}{\omega+\omega_{\lambda}}\sum_{m} g_{\lambda
m}\left[\frac{b_{m}(0)}{\omega-\Omega_{m}}+
\frac{b_{m}^{\dag}(0)}{\omega+\Omega_{m}}\right]   -
\frac{\hbar^{-2}}{\omega+\omega_{\lambda}}
\sum_{m,\lambda'}\frac{2 g_{\lambda
m}\Omega_{m}}{\omega^{2}-\Omega_{m}^{2}} \left[ g_{\lambda'
m}\;a_{\lambda'}(\omega) + g_{\lambda'
m}^{*}\;a^{\dag}_{\lambda'}(\omega) \right].
\label{fieldcreafterintegratingout}
\end{eqnarray}
\end{subequations}
\end{widetext}
The optical modes are no longer independent because of the
interaction with the atoms. The three terms in the right-hand
sides of Eqs.~(\ref{fieldannafterintegratingout}) and
(\ref{fieldcreafterintegratingout}) can be related to three
reasons why there can be light in mode $\lambda$: firstly, because
there is light in the undisturbed mode that has not `seen' the
atom; secondly, because the atom can emit light into the mode
$\lambda$; the third term describes  transitions of light in and
out of the mode $\lambda$ to and from modes $\lambda'$, due to
scattering off one of the guest atoms. Since the
relations~(\ref{fieldoperatorsafterintegratingout}) are implicit
rather than explicit solutions for the operators, the
identification of terms in the equations with scattering and
emission processes can only be approximate.

The results~(\ref{fieldoperatorsafterintegratingout}) for the
creation and annihilation operators can be directly used with
Eq.~(\ref{ddooreps}) to find the following equation for the field
${\bf F}$
\begin{subequations}\label{eqforfieldexactlipp}
\begin{eqnarray}
{\bf F}({\bf r},\omega) & = & {\bf E}^{(0)}({\bf r},\omega) \label{undisturbed} \\
& + & \sum_{m} \bfsfK({\bf r},{\bf R}_{m},\omega)\cdot {\bf
S}_{m}(\omega) \label{sourceterm} \\ & + & \sum_{m}\bfsfK({\bf
r},{\bf R}_{m},\omega)\cdot \bfsfV_{m}(\omega)\cdot {\bf F}({\bf
R}_{m},\omega). \label{scatteringterm}
\end{eqnarray}
\end{subequations}
This is the central result of this paper. It is an exact
Lippmann-Schwinger equation and it describes the resonant
scattering off and emission by guest atoms inside an inhomogeneous
dielectric, both for strong and for weak atom-field interactions.
The equation has an undisturbed term~(\ref{undisturbed}), a source
term~(\ref{sourceterm}), and a scattering
term~(\ref{scatteringterm}).

The elements of Eq. (\ref{eqforfieldexactlipp}) must still be
explained. The operator ${\bf E}^{(0)}({\bf r},\omega)$ is the
electric field in the absence of the atoms, with both the positive
and negative frequency parts. The atomic source operators ${\bf
S}_{m}(\omega)$ are vectors that have the form $\hat{\bm
\mu}_{m}S_{m}(\omega)$, where ${\hat{\bm \mu}}_{m}$ denotes the
unit vector in the direction of the atomic dipole moment ${\bm
\mu}_{m}$ and
\begin{equation}\label{atomicsourceoperator}
S_{m}(\omega)\equiv \left(\frac{-i
\mu_{m}\omega^{2}}{\varepsilon_{0}c^{2}}\right)\left[
\frac{b_{m}(0)}{\omega-\Omega_{m}}+
\frac{b_{m}^{\dag}(0)}{\omega+\Omega_{m}}\right].
\end{equation}
 Notice that ${\bf S}_{m}$ features the atomic creation and
annihilation operators at the initial time zero: in quantum
optics, the atomic variables can not be completely integrated out
in an `all-light' picture.

The optical potentials $\bfsfV_{m}(\omega)$ produced by the atoms
are dyadics equal to $\hat{{\bm \mu}}_{m}V_{m}(\omega)\hat{{\bm
\mu}}_{m}$, where
\begin{equation}\label{Vmnorwa}
V_{m}(\omega) \equiv
\left(\frac{\mu_{m}^{2}\omega^{2}}{\hbar\varepsilon_{0}c^{2}}\right)
\left(\frac{2\Omega_{m}}{\omega^{2}-\Omega_{m}^{2}}\right).
\end{equation}
Both the sources and the potentials have resonances at frequencies
$\pm \Omega_{m}$. Potentials $V_{m}(\omega)$ are sometimes
rewritten as $-(\omega/c)^{2}$ times a `bare polarizability'
$\alpha_{{\rm B}m}(\omega)$ \cite{DeVries98a}. In the present
case, the bare polarizabilities are real (except exactly on
resonance) and they change sign when going through their
resonances at $\Omega_{m}$; the resonances are infinitely sharp
because all possible nonradiative decay processes are neglected;
the polarizability is called `bare'  because it does not (and
should not)  contain radiative broadening of its resonance (but
see Sec.~\ref{Singlepoint}).

The last undefined factor in Eq.~(\ref{eqforfieldexactlipp}) is
the dyadic quantity $\bfsfK$ which is given by
\begin{equation}\label{Kdefgreen}
\bfsfK({\bf r},{\bf r'},\omega) \equiv c^{2}\sum_{\lambda}
\frac{{\bf f}_{\lambda}({\bf r}){\bf f}_{\lambda}^{*}({\bf
r'})}{(\omega^{2}-\omega_{\lambda}^{2})}\cdot\frac{\omega_{\lambda}^{2}}{\omega^{2}}
\end{equation}
Usually, in a Lippmann-Schwinger equation one finds the Green
function (called $\bfsfG$) of a medium where we now find the
dyadic $\bfsfK$. Interestingly, $\bfsfK$ turns out to be different
from $\bfsfG$, even for free space, as will be studied in
Sec.~\ref{Greenmedium}. All the elements of
Eq.~(\ref{eqforfieldexactlipp}) have now been defined.

 Another important
field operator for the medium is the vector potential ${\bf A}$.
The magnetic field ${\bf B}$ equals $\nabla\times {\bf A}$. In the
canonical quantization theories \cite{Dalton,Wubs03} upon which
our Hamiltonian~(\ref{Hafquantcorrglobal}) is based, the
generalized Coulomb gauge is chosen, which means that ${\bf A}$ is
generalized transverse. Its expansion in terms of the normal modes
is given below.  With
Eq.~(\ref{fieldoperatorsafterintegratingout}) this leads to
\begin{subequations}\label{vectorpotintermsoffieldall}
\begin{eqnarray}\label{vectorpotintermsoffield_a}
{\bf A}({\bf r},\omega) & \equiv &
 \sum_{\lambda}\sqrt{\frac{\hbar}{2\varepsilon_{0}\omega_{\lambda}}}\left[
a_{\lambda}(\omega)\;{\bf f}_{\lambda}({\bf r}) +
a_{\lambda}^{\dag}(\omega)\;{\bf f}_{\lambda}^{*}({\bf
r})\;\right]  \\
& = & {\bf A}^{(0)}({\bf r},\omega) \nonumber \\ & + &
\frac{1}{i\omega}\sum_{m}\bfsfG^{\rm T}({\bf r}, {\bf
R}_{m},\omega)\cdot{\bf S}_{m}(\omega) \nonumber \\ &+&
 \frac{1}{i\omega}\sum_{m}\bfsfG^{\rm T}({\bf r}, {\bf
R}_{m},\omega)\cdot\bfsfV_{m}(\omega)\cdot{\bf F}({\bf R}_{m},
\omega). \label{vectorpotintermsoffield_b}
\end{eqnarray}
\end{subequations}
Analogously to Eq.~(\ref{eqforfieldexactlipp}), an undisturbed
term, a source term and a scattering term can be identified for
the vector potential.

A difference between Eq.~(\ref{eqforfieldexactlipp}) for the field
${\bf F}$ and Eq.~(\ref{vectorpotintermsoffield_b}) for ${\bf A}$
is that only the former is a Lippmann-Schwinger equation and that
${\bf A}$ immediately follows from the solution of ${\bf F}$,
rather than {\em vice versa}.  In a minimal-coupling formalism,
one would find a Lippmann-Schwinger equation for the vector
potential instead. Another important difference between the
equations for the two fields is that in
Eq.~(\ref{vectorpotintermsoffield_b}) for {\bf A} the generalized
transverse Green function $\bfsfG^{\rm T}$ appears, rather than
the dyadic $\bfsfK$ of Eq.~(\ref{eqforfieldexactlipp}).
Definitions of and relations between $\bfsfG$, $\bfsfG^{\rm T}$,
and $\bfsfK$ will be studied shortly, in Sec.~\ref{Greenmedium}.

Often, Lippmann-Schwinger equations are derived in `all-light'
formalisms that
  start with a given optical  potential as a perturbation.
   Here instead, the approach started one level deeper and the optical potential
   $\bfsfV_{m}$ is output rather than input. An important
feature  in Eq.~(\ref{eqforfieldexactlipp}) is that the atoms are
not only point scatterers (potentials), but also point sources for
light. Both appear as two sides of the same coin in one equation.
Solutions for the equation will be discussed shortly  in
Sec.~\ref{Singlepoint} for one atom and in
Sec.~\ref{severalsources} for several atoms.

\section{Green functions of the medium}\label{Greenmedium}

The dyadic quantities $\bfsfK$ and $\bfsfG^{\rm T}$ will now be
related to the Green function of the medium.  The  (full) Green
tensor $\bfsfG({\bf r},{\bf r'},\omega)$ of an inhomogeneous
medium characterized by the dielectric function $\varepsilon({\bf
r})$ is the solution of the wave equation
\begin{equation}\label{Greenmediumfull}
-{\bm \nabla}\times{\bm \nabla}\times\bfsfG({\bf r},{\bf r'},
\omega) + \varepsilon({\bf r})(\omega/c)^{2}\bfsfG({\bf r},{\bf
r'}, \omega) = \delta({\bf r}-{\bf r'})\bfsfI,
\end{equation}
where the right-hand side is the ordinary Dirac delta function
times the unit tensor.

For a discussion of $\bfsfG^{\rm T}$, it is useful to first
introduce the concept of a generalized transverse delta function
\cite{Glauber91,Wubs03}. (For comparison, Green and delta
functions of a homogeneous medium are given in the Appendix.) A
generalized transverse delta function ${\bm
\delta}_{\varepsilon}^{\rm T}$ (a distribution) can be defined in
terms of the mode functions ${\bf f}_{\lambda}$ [see
Eq.~(\ref{true modes})]:
\begin{equation}\label{deltagegtrans}
{\bm \delta}_{\varepsilon}^{\rm T}({\bf r},{\bf r'}) \equiv
\sum_{\lambda} {\bf f}_{\lambda}^{*}({\bf r}){\bf
f}_{\lambda}({\bf r'})\varepsilon({\bf r'}).
\end{equation}
Now ${\bm \delta}_{\varepsilon}^{\rm T}$ has the projection
property $\int\mbox{d}{\bf r}_{1}\;\bar{\bm
\delta}_{\varepsilon}^{\rm T}({\bf r}_{1},{\bf r})\cdot{\bf
X}^{\rm T}({\bf r}_{1})
  = {\bf X}^{\rm T}({\bf r})$ for all (ordinary) transverse vector
  fields ${\bf X}^{\rm T}$. The bar in $\bar{{\bm \delta}}_{\varepsilon}^{\rm T}$ denotes the
transpose. The same projection can be applied to
  Eq.~(\ref{Greenmediumfull}). In doing so,
 the transverse double-curl term is projected onto
itself. The generalized transverse Green function ${\bfsfG}^{\rm
T}$ can now be defined such that $\varepsilon({\bf r})\bfsfG^{\rm
T} ({\bf r},{\bf r}',\omega)$ equals the projection
$\int\mbox{d}{\bf r}_{1}\bar{\delta}_{\varepsilon}^{\rm T}({\bf
r}_{1},{\bf r})\cdot\left[\varepsilon({\bf r}_{1})\bfsfG({\bf
r}_{1}, {\bf r}')\right]$. The projection then leads to the
following equation for $\bfsfG^{\rm T}$:
\begin{equation}\label{Greengentrans}
-{\bm \nabla}\times{\bm \nabla}\times\bfsfG({\bf r},{\bf r'},
\omega) + \varepsilon({\bf r})(\omega/c)^{2}\bfsfG^{\rm T}({\bf
r},{\bf r'}, \omega) = \bar{{\bm \delta}}_{\varepsilon}^{\rm
T}({\bf r'},{\bf r}).
\end{equation}
Notice that $\bfsfG$ rather than $\bfsfG^{\rm T}$ appears in the
first term. Furthermore, a longitudinal Green function
$\bfsfG^{\rm L}$ can be defined as $\bfsfG-\bfsfG^{\rm T}$. By
taking the difference of Eq.~(\ref{Greenmediumfull}) and
Eq.~(\ref{Greengentrans}) one can see that $\bfsfG^{\rm L}$ has
the form
\begin{subequations}\label{Greengenlong}
\begin{eqnarray}
\bfsfG^{\rm L}({\bf r},{\bf r'}) & \equiv &
\frac{1}{\varepsilon({\bf r})(\omega/c)^{2}}\left[\delta({\bf
r}-{\bf r'})\bfsfI-\bar{{\bm \delta}}_{\varepsilon}^{\rm T}({\bf
r'},{\bf r})\right] \label{GLdef} \\ & \equiv &
\frac{1}{\varepsilon({\bf r})(\omega/c)^{2}}\bar{{\bm
\delta}}_{\varepsilon}^{\rm L}({\bf r'},{\bf r}),
\label{deltaLdef}
\end{eqnarray}
\end{subequations}
In equality  (\ref{deltaLdef}) the generalized longitudinal delta
function $\delta_{\varepsilon}^{\rm L}$ was defined as the
difference between the ordinary Dirac and the generalized
transverse delta function,  so that ${\bm
\delta}_{\varepsilon}^{\rm T}+{\bm \delta}_{\varepsilon}^{\rm
L}=\delta\bfsfI$. We called $\bfsfG^{\rm L}$ the longitudinal
Green function, but it is not self-evident that for every
inhomogeneous dielectric $\bfsfG^{\rm L}$ is longitudinal indeed.
Proofs that $\int\mbox{d}{\bf r'} \bfsfG^{\rm L}({\bf r}, {\bf
r'})\cdot{\bf X}^{\rm T}({\bf r'})=0$ and also that
$\int\mbox{d}{\bf r} {\bf X}^{\rm T}({\bf r})\cdot \bfsfG^{\rm
L}({\bf r}, {\bf r'})=0$ can be found with the help of Eqs.~(32a)
and (32b) of Ref.~\cite{Wubs03}, respectively.  Then, since
$\bfsfG^{\rm L}$ is longitudinal, $\bfsfG$ in
Eq.~(\ref{Greengentrans}) can be replaced by $\bfsfG^{\rm T}$.
Hence the projection of Eq.~(\ref{Greenmediumfull}) leads to a
unique defining equation for $\bfsfG^{\rm T}$.

 From
Eqs.~(\ref{true modes}) and (\ref{Greengentrans}), it follows that
the generalized transverse Green tensor $\bfsfG^{\rm T}$ has the
mode expansion
\begin{equation}\label{gefieldzonderatoom}
\bfsfG^{\rm T}({\bf r},{\bf r}',\omega) =
c^{2}\sum_{\lambda}\frac{{\bf f}_{\lambda}({\bf r})\;{\bf
f}_{\lambda}^{*}({\bf
r'})}{(\omega+i\eta)^{2}-\omega_{\lambda}^{2}}.
\end{equation}
In this manifestly generalized transverse form, $\bfsfG^{\rm T}$
appeared in Eq.~(\ref{vectorpotintermsoffield_b}) for the vector
potential. In the denominator of Eq.~(\ref{gefieldzonderatoom}) we
have for once  made explicit the positive and infinitesimally
small imaginary part of the frequency $\omega$, through the term
$i \eta$. With the positive sign of the imaginary part,
(\ref{gefieldzonderatoom}) is the causal Green function which
transformed back to the time-domain gives a Green function
$\bfsfG^{\rm T}({\bf r},{\bf r}',t-t_{0})$ which is nonzero only
for positive time differences $(t-t_{0})$.

We are now in the position to rewrite and interpret the dyadic
$\bfsfK$ [see Eq.~(\ref{Kdefgreen})] that appears in the
Lippmann-Schwinger equation (\ref{eqforfieldexactlipp}) for the
field ${\bf F}$:
\begin{eqnarray}\label{Kredef}
\bfsfK({\bf r},{\bf r'},\omega) & = &  c^{2}\sum_{\lambda}
\frac{{\bf f}_{\lambda}({\bf r}){\bf f}_{\lambda}^{*}({\bf
r'})}{\omega^{2}-\omega_{\lambda}^{2}} -
(c/\omega)^{2}\sum_{\lambda}{\bf f}_{\lambda}({\bf r}){\bf
f}_{\lambda}^{*}({\bf r'}) \nonumber  \\
& = & \bfsfG^{\rm T}({\bf r},{\bf r'},\omega) -
\frac{1}{\varepsilon({\bf r})(\omega/c)^{2}}\;\bar{\bm
\delta}_{\varepsilon}^{\rm T}({\bf r'},{\bf r}).
\label{Kredefsecondline}
\end{eqnarray}
 It consists of the generalized transverse Green
function (\ref{gefieldzonderatoom}) and a term proportional to the
transpose of the generalized transverse delta function, $\bar{\bm
\delta}_{\varepsilon}^{\rm T}$ (\ref{deltagegtrans}).  Both terms
 are medium-dependent. Note that $\bfsfK$ is generalized
transverse in its variable ${\bf r}$. If only because of this
 property,  $\bfsfK$ is not equal to the
total Green function (\ref{Greenmediumfull}). Nevertheless, the
definition (\ref{deltaLdef}) of the longitudinal Green function
can be used to rewrite $\bfsfK$ as
\begin{equation}\label{Kredef2}
\bfsfK({\bf r},{\bf r'},\omega) = \bfsfG({\bf r},{\bf r'},\omega)
- \frac{1}{\varepsilon({\bf r})(\omega/c)^{2}}\delta({\bf r}-{\bf
r'})\bfsfI.
\end{equation}
According to this identity, the dyadic $\bfsfK$   differs from the
full Green function of the medium only when its two position
arguments ${\bf r}$ and ${\bf r'}$ coincide. Although different
from $\bfsfG$, the quantity $\bfsfK$ will also be called a Green
function.  The occurrence of $\bfsfK$ rather than $\bfsfG$ in the
Lippmann-Schwinger equation will be discussed further in the
Appendix, where the volume-integrated electric field around an
atom is calculated.

\section{Single-atom properties altered by the
medium}\label{Singlepoint} \subsection{Solution of the LS
equation} An atom in a group of atoms in an inhomogeneous
dielectric will have different properties as compared to free
space, because of the dielectric and because of the other atoms.
In this section the effect of the medium on the individual atoms
will be  considered. The next and major step, in section
\ref{severalsources}, will be to study some effects that the
medium-modified atoms can have on each other.

Assume that in the dielectric there is only one guest atom present
with dipole moment ${\bm \mu}$ and transition frequency $\Omega$.
The effect of the medium on the scattering and emission properties
of the atom can be found by solving
Eq.~(\ref{eqforfieldexactlipp}) exactly  by successive iterations
\begin{eqnarray}\label{fieldthroughiteration}
{\bf F} & = &\left[{\bf E}^{(0)} + \bfsfK\cdot{\bf S}\right]
\nonumber \\& + & \bfsfK\cdot \bfsfV\cdot \left[{\bf E}^{(0)} +
\bfsfK\cdot{\bf S}\right] \nonumber \\ & + & \bfsfK\cdot
\bfsfV\cdot\bfsfK\cdot \bfsfV\cdot \left[{\bf E}^{(0)} +
\bfsfK\cdot{\bf S}\right] +\ldots
\end{eqnarray}
In this equation, $\bfsfK$ and $\bfsfV$ are classical quantities,
whereas ${\bf F}$, ${\bf E}^{(0)}$, and ${\bf S}$ are quantum
mechanical operators. The infinite series of multiple-scattering
terms can be summed to give
\begin{equation}\label{Ftotsolved}
{\bf F}({\bf r}, \omega)  =  {\bf E}^{(0)}({\bf r},\omega) +{\bf
F}_{\rm scat}({\bf r},\omega) + {\bf F}_{\rm source}({\bf
r},\omega),
\end{equation}
where, as before, ${\bf E}^{(0)}({\bf r},\omega)$ is the
electric-field operator of the inhomogeneous medium in the absence
of the guest atoms.

The operator ${\bf F}_{\rm scat}({\bf r},\omega)$ in
Eq.~(\ref{Ftotsolved}) describes light that is scattered by the
guest atom and it has the form
\begin{equation}\label{Fscattdef}
{\bf F}_{\rm scat}({\bf r},\omega)  =
 \bfsfK({\bf r},{\bf R},\omega)\cdot \bfsfT(\omega)\cdot
{\bf E}^{(0)}({\bf R},\omega),
\end{equation}
with the single-atom T-matrix  defined by
\begin{equation}\label{tatominhomogene}
\bfsfT(\omega) = {\hat {\bm \mu}} T(\omega) {\hat {\bm \mu}} =
{\hat {\bm \mu}} \left[\frac{V(\omega)}{1 - {\hat {\bm \mu}}\cdot
\bfsfK({\bf R},{\bf R},\omega)\cdot {\hat {\bm \mu}}
V(\omega)}\right]{\hat {\bm \mu}}.
\end{equation}
The T-matrix  is sometimes written as $-(\omega/c)^{2}$ times a
dynamical polarizability $\alpha(\omega)$ [compare with
Eq.~(\ref{Vmnorwa})] and both depend on the atomic position inside
the inhomogeneous dielectric. The expectation value of the
scattered field (\ref{Fscattdef}) only depends on the initial
quantum state of the light (through the term ${\bf E}^{(0)}$).
Unlike for a two-level atom, the light-scattering properties of a
harmonic-oscillator atom do not depend on the atomic excitation.
The scattering process can be read from right to left in the
right-hand side of Eq.~(\ref{Fscattdef}): light ${\bf E}^{(0)}$
that has not yet seen the atom scatters off the atom (as described
by $\bfsfT$), and the scattered part of the light propagates
through the dielectric as described by $\bfsfK$.

Finally, there is in Eq.~(\ref{Ftotsolved}) the source field
\begin{eqnarray}\label{sourcefielddef}
{\bf F}_{\rm source}({\bf r},\omega) & = &  \bfsfK({\bf r},{\bf
R},\omega)\cdot{\bf S}(\omega)  \\ & + & \bfsfK({\bf r},{\bf
R},\omega)\cdot \bfsfT(\omega)\cdot \bfsfK({\bf R},{\bf
R},\omega)\cdot{\bf S}(\omega). \nonumber
\end{eqnarray}
Expectation values of the source field ${\bf F}_{\rm source}$ only
depend on the initial atomic state. Notice that the same T-matrix
that shows up in the scattered field~(\ref{Fscattdef}) also
appears in the source field~(\ref{sourcefielddef}). Light emitted
by an atomic point source will be studied further in
Sec.~\ref{solscatsource} and scattered light in
Sec.~\ref{pointscatsolution}.

\subsection{Light emitted by a point source}
\label{solscatsource}

The source
field~(\ref{sourcefielddef}) can be rewritten as
\begin{equation}\label{F2}
{\bf F}_{\rm source}({\bf r},\omega) = \frac{\bfsfK({\bf r},{\bf
R},\omega)\cdot {\bf S}(\omega)}{1 - {\hat{\bm
\mu}}\cdot\bfsfK({\bf R},{\bf R},\omega)\cdot {\hat{\bm \mu}}
V(\omega)},
\end{equation}
with ${\bf S}(\omega)$ as defined in
Eq.~(\ref{atomicsourceoperator}). The time dependence of the
source field at the position ${\bf r}$ due to the presence of the
source at ${\bf R}$ follows from the inverse Laplace transform of
Eq.~(\ref{F2}),
\begin{equation}\label{f2rt}
{\bf F}_{\rm source}({\bf r},t)=
\frac{1}{2\pi}\int_{-\infty}^{\infty} \mbox{d}\omega\;e^{-i\omega
t} {\bf F}_{\rm source}({\bf r},\omega).
\end{equation}
 This integral  can not be evaluated further without the explicit
knowledge of the
 Green function $\bfsfK$. The source field decays in time due to spontaneous
emission by the atom. The decay rate  can be found by multiplying
numerator and denominator in Eq.~(\ref{F2}) by
$(\omega^{2}-\Omega^{2})$ and by realizing that the zeroes $\omega
= \Omega_{1}(\Omega)$ of $\omega^{2}-\Omega^{2} -2\Omega
X(\omega)$ with
\begin{equation}\label{Xdefforoneatom}
X(\omega) \equiv \hat{\bfmu}\cdot\bfsfK({\bf R},{\bf
R},\omega)\cdot \hat{\bfmu}
\left[\mu^{2}\omega^{2}/(\hbar\varepsilon_{0}c^{2})\right]
\end{equation}
are the frequency poles of the source field. Until now, our
solution is exact. At this point we make a pole approximation,
which is only valid if the atom-field coupling is weak.  The pole
approximation gives $\Omega_{1}= \Omega + X(\Omega)$, with
$X(\Omega)$ the difference between the dressed resonance frequency
$\Omega_{1}(\Omega)$ and the bare atomic resonance frequency
$\Omega$. The exponential (amplitude) spontaneous-decay rate is
\begin{equation}\label{decaysinglesource}
\Gamma/2 \equiv -  \mbox{Im}\;X(\Omega).
\end{equation} The
decay rate of the intensity of the field is $\Gamma$. It is
nonnegative by definition of $\bfsfK$ in Eq.~(\ref{Kdefgreen}).
The delta function term in $\bfsfK$ (\ref{Kredef2}) and the
longitudinal Green function  $\bfsfG^{\rm L}({\bf R},{\bf
R},\Omega)$ in Eq.~(\ref{GLdef}) are real quantities, so that
$\Gamma$ is proportional to the imaginary part of only the
generalized transverse Green function $\bfsfG^{\rm T}$. The
pro\-per\-ty that
 $\bfsfG^{\rm L}$ does not
contribute to the spontaneous-emission rate is a generalization of
the well-known result for homogeneous dielectrics \cite{Barnett96}
and it only holds for non-absorbing dielectrics. Using the mode
composition Eq.~(\ref{gefieldzonderatoom}) of $\bfsfG^{\rm T}$, we
find $\Gamma =
\pi/(\hbar\varepsilon_{0})\sum_{\lambda}|\bfmu\cdot{\bf
f}_{\lambda}({\bf
r})|^{2}\omega_{\lambda}\delta(\Omega-\omega_{\lambda})$, the same
expression that one also finds from Fermi's golden rule
\cite{Glauber91}.  The decay rate depends both on the atom's
position and on its orientation inside the inhomogeneous
dielectric. For free space, the imaginary part of $\bfsfG_{0}^{\rm
T}({\bf R},{\bf R},\Omega)$ is equal to $-\Omega/(6\pi c)\bfsfI$
[see Eq.~(\ref{g0transpoint})]. This gives the familiar free-space
spontaneous-decay rate $\Gamma_{0} = \mu^{2}\Omega^{3}/(3 \pi
\hbar \varepsilon_{0} c^{3})$.

The dressed resonance frequency $\Omega_{1}$ can be written as
$\Omega + \Delta'(\Omega) - i \Gamma(\Omega)/2$. Apart from a
decay rate there is a frequency shift $\Delta'$ that is equal to
$\mbox{Re}\; X(\Omega)$. For two reasons, $\Delta'$ is infinitely
large even for
 free space. Firstly, the delta function term
$\delta({\bf r}-{\bf R})\bfsfI /[\varepsilon({\bf
R})(\omega/c)^{2}]$ in Eq.~(\ref{Kredef2}) diverges when ${\bf r}$
and ${\bf R}$ are equal. This self-interaction term is
medium-dependent through the factor $\varepsilon({\bf R})$, but
here and in the following we assume that guest atoms are
electronically well separated from the dielectric medium, so that
an empty-cavity model applies where the relative dielectric
function is equal to unity at the position of the guest atom
\cite{Wubs03}. The second reason why $\Delta'$ diverges is well
known for free space: $\bfsfG_{0}({\bf r},{\bf R},\omega)$
diverges when ${\bf r}$ approaches ${\bf R}$ [see
Eq.~(\ref{GTL})]. By a procedure called mass renormalization, the
combined radiative shift in  free space becomes finite, see for
example \cite{Milonni94}.  From now on we can assume that $\Omega$
is the observable atomic frequency in free space; inside a
dielectric, the atomic frequency shifts by an amount $\Delta$ that
is given by the real part of $[X(\omega) - X_{0}(\Omega)]$, or in
terms of the Green functions
\begin{equation}\label{reallevelshift}
\Delta = {\hat{\bm \mu}}\cdot\mbox{Re} \left[\;\bfsfG({\bf R},{\bf
R},\Omega)-\bfsfG_{0}({\bf R},{\bf R},\Omega)\;\right]\cdot {\hat
{\bm \mu}}
\left(\frac{\mu^{2}\Omega^{2}}{\hbar\varepsilon_{0}c^{2}}\right).
\end{equation}
The shift  depends on the atomic position and dipole orientation.
Notice that the full Green function is needed to determine the
line shift, whereas for the decay rate it sufficed to know
$\bfsfG^{\rm T}$.

The position-dependent radiative shifts are a mechanism of
inhomogeneous broadening of the detected light.  Electronic shifts
usually dominate inhomogeneous broadening. Experimentally it will
be hard to single out radiative shifts (a photonic effect) from
electronic line shifts (due to changes in the atomic wave
functions inside the medium).

\subsection{Light scattered by a point scatterer}\label{pointscatsolution}
In  the scattered field of Eq.~(\ref{Fscattdef}), the atom appears
as a point scatterer with an internal resonance in the optical
potential $V(\omega)$ and a corresponding resonance in the
T-matrix in  Eq.~(\ref{tatominhomogene}).  The scattered field
 has frequency poles in the T-matrix (just like the source field), but it also
 has  poles for every optical mode frequency  $\pm \omega_{\lambda}$ (unlike the source field).
 The time-dependence of the scattered field can be
understood by separating the frequency poles (straightforward, but
not spelled out here), again followed by an inverse Laplace
transformation. In the following, place the atom in the origin.
For the part of ${\bf F}_{\rm scat}$ featuring the annihilation
operators, one finds
\begin{eqnarray}\label{F1plussolved}
&&{\bf F}_{\rm scat}^{(+)}({\bf r}, t)  =   \sum_{\lambda}
\frac{-\mu^{2}}{2\pi\hbar\varepsilon_{0}c^{2}}
\sqrt{\frac{\hbar\omega_{\lambda}}{2\varepsilon_{0}}}
a_{\lambda}^{(0)}(0){\bf f}_{\lambda}({\bf 0})\cdot \hat{\bm
\mu}\times \nonumber
\\ && \int_{-\infty}^{\infty}\mbox{d}\omega \omega^{2} e^{-i\omega t}
  \bfsfK({\bf r},{\bf 0}, \omega)\cdot\hat{\bm \mu}\times \nonumber \\  \biggl\{ & - &\frac{2\Omega}{(\Omega+\Delta)^{2}
-\omega_{\lambda}^{2}+\Gamma^{2}/4-i\omega_{\lambda}\Gamma} \cdot
\frac{1}{\omega-\omega_{\lambda}} \nonumber \\ &  + &
\frac{\Omega/(\Omega+\Delta)}{\Omega+\Delta
-\omega_{\lambda}-i\Gamma/2}\cdot
\frac{1}{\omega-\Omega -\Delta +i \Gamma /2} \nonumber \\
& + & \frac{\Omega/(\Omega+\Delta)}{\Omega+\Delta
+\omega_{\lambda}+i\Gamma/2}\cdot
\frac{1}{\omega+\Omega+\Delta+i\Gamma/2}
 \;\;\biggl\}.
\end{eqnarray}
The negative-frequency part ${\bf F}^{(-)}$ of the field  equals
$[{\bf F}^{(+)}]^{\dag}$. The three terms between curly brackets
in (\ref{F1plussolved}) cor\-res\-pond to different optical
processes. The first term describes elastic light scattering by
the guest atom inside the inhomogeneous dielectric; the second
term has an exponentially decaying time dependence and corresponds
to resonance fluorescence; finally, the third term is an
exponentially decaying nonresonant term, corresponding to an
utterly improbable process that one could call anti-resonance
fluorescence. In a rotating-wave approximation this process would
disappear. After neglecting this third term, all the
$\omega$-poles in the integral (\ref{F1plussolved}) have positive
real parts, so that (\ref{F1plussolved}) can be called the
positive-frequency part of the field ${\bf F}_{\rm scat}$.

 Now consider the second term
in Eq.~(\ref{F1plussolved}) in more detail. In the resonance
fluorescence process the guest atom is excited by light of
frequency $\omega_{\rm p}$, after which the atomic source decays
exponentially due to light emission at frequency $\Omega_{s}$. (In
contrast, elastically scattered light oscillates with the pump
frequency $\omega_{p}$.) The fluorescent light has the same
position-dependent emission rates $\Gamma({\bf R},\Omega)$
(\ref{decaysinglesource}) and line shifts $\Delta({\bf R},\Omega)$
(\ref{reallevelshift}) as found for spontaneous emission before. A
difference between the source-field of Eq.~(\ref{F2}) and the
fluorescent light in Eq.~(\ref{F1plussolved}) is that the latter
also contains the information how well the pump light that comes
in via mode $\lambda=p$ can excite the atom, in the factor ${\bf
f}_{p}(0)\cdot \hat{\bm \mu}$. This difference is especially
important for inhomogeneous dielectrics, where  atoms will be
excited easier here than there. And indeed, it is through the
process of resonance fluorescence that lifetimes of atoms in
dielectric media are usually measured.

In a resonance fluorescence experiment, a light pulse or wave
packet passes the atom during a time $T$. In
expression~(\ref{F1plussolved}), the intensity of resonantly
emitted light depends on the expectation value with respect to the
quantum state of light at time $t=0$. This can only be a valid
description of the process if $T\ll \Gamma^{-1}$, in other words,
if the wave packet is so short that it ``prepares percussionally
the excited state'' (\cite{Cohen92}, p.~97) of the atom at time
$t=0$. This is typically the case, even if the medium broadens the
excitation pulse: excitation pulses last picoseconds and lifetimes
lie in the nanosecond regime.

\section{Several atoms as point sources and
scatterers}\label{severalsources}

\subsection{Solution of the LS equation}\label{solLSSeveralatoms}

In section~\ref{Singlepoint} it was found how scattering by and
emission rates of single atoms are  influenced by their dielectric
surroundings. In the present section it is studied how the
medium-modified atoms can influence each other. The atomic wave
functions are assumed not to overlap each other and to be
unaffected by the dielectric. The atomic positions can be
arbitrary, so we can decide to choose the atoms on a line
\cite{Clemens03} or on a lattice \cite{Nienhuis87} or at random
positions. The general method to solve the Lippmann-Schwinger
equation~(\ref{eqforfieldexactlipp}) in this more complicated
situation is outlined here. In Sec.~\ref{supertwo}, the formalism
will be used to study two-atom superradiance inside an
inhomogeneous dielectric medium.

For one atom, the Lippmann-Schwinger
equation~(\ref{eqforfieldexactlipp}) was solved by summing a
series to all orders of the atomic potential. In the present
many-atom case all atomic transition dipole moments, orientations,
and frequencies are allowed to be different so that also all
optical potentials $\bfsfV_{j}$ are different. The LS equation
will now be solved by efficiently summing a somewhat more
complicated series.  Use the abbreviations ${\bf F}={\bf F}({\bf
r}, \omega)$, ${\bf F}_{m}={\bf F}({\bf R}_{m}, \omega)$,
$\bfsfK_{m} = \bfsfK({\bf r},{\bf R}_{m},\omega)$, $\bfsfK_{mn} =
\bfsfK({\bf R}_{m},{\bf R}_{n},\omega)$, and introduce ${\bf
F}^{(1)} \equiv {\bf E}^{(0)}({\bf r},\omega) + \sum_{m}
\bfsfK({\bf r},{\bf R}_{m},\omega)\cdot {\bf S}_{m}(\omega)$.
Also, ${\bf F}^{(1)}_{n}$ is shorthand for ${\bf F}^{(1)}({\bf
R}_{n})$. By iteration it follows that  the field $({\bf F}-{\bf
F}^{(1)})$ of Eq.~(\ref{eqforfieldexactlipp}) becomes
\begin{subequations}
\begin{eqnarray}\label{solveseveralatomLS}
&&    \sum_{n=1}^{N}\bfsfK_{n}\cdot \bfsfV_{n}\cdot {\bf
F}^{(1)}_{n} + \sum_{m,n=1}^{N}\bfsfK_{m}\cdot \bfsfV_{m}\cdot
\bfsfK_{mn} \cdot \bfsfV_{n} \cdot {\bf F}^{(1)}_{n} \nonumber \\
&& +  \sum_{m,p,n=1}^{N}\bfsfK_{m}\cdot \bfsfV_{m}\cdot
\bfsfK_{mp} \cdot\bfsfV_{p}\cdot \bfsfK_{pn} \cdot \bfsfV_{n}
\cdot {\bf F}^{(1)}_{n} +\ldots
 \label{solveseveralatomLSinV}
\end{eqnarray}
This  can conveniently be rewritten in terms of the single-atom
T-matrices of Eq.~(\ref{tatominhomogene}) as
\begin{eqnarray}\label{solveseveralatomLSinT}
&&  \sum_{n=1}^{N}\bfsfK_{n}\cdot \bfsfT_{n}\cdot {\bf
F}^{(1)}_{n} + \sum_{m,n=1}^{N}\bfsfK_{m}\cdot \bfsfT_{m}\cdot
\bfsfK'_{mn} \cdot \bfsfT_{n} \cdot {\bf F}^{(1)}_{n} \nonumber
\\&& + \sum_{m,p,n=1}^{N}\bfsfK_{m}\cdot
\bfsfT_{m}\cdot \bfsfK'_{mp} \cdot \bfsfT_{p} \cdot \bfsfK'_{pn}
\cdot \bfsfT_{n}\cdot {\bf F}^{(1)}_{n} +\ldots.
\end{eqnarray}
\end{subequations}
Here the tensor $\bfsfK'_{mn}$ is defined as
$(1-\delta_{mn})\bfsfK_{mn}$, which by virtue of
Eq.~(\ref{Kredef2}) is equal to $\bfsfG'_{mn}\equiv
(1-\delta_{mn})\bfsfG_{mn}$. A single-atom T-matrix already sums
up all multiple potential-scattering off a single atom, which
explains that neighboring T-matrices in terms of this series
belong to different atoms. The equivalence of
Eqs.~(\ref{solveseveralatomLSinV}) and
(\ref{solveseveralatomLSinT}) can be seen by expanding single-atom
T-matrices in terms  of single-atom potentials. Now every
higher-order term in Eq.~(\ref{solveseveralatomLSinT}) can be
constructed from the previous-order term by inserting into the
latter the $N\times N$ matrix with $(i,j)$-elements ${\hat {\bm
\mu}}_{i}\cdot\bfsfG'_{ij} \cdot \bfsfT_{j}\cdot {\hat {\bm
\mu}}_{j}$. By summing the geometric series of matrices and
dropping the abbreviations, it follows that
\begin{eqnarray}\label{TNexactsum}
{\bf F}({\bf r}, \omega) &  = & {\bf F}^{(1)}({\bf r}, \omega) \\
& + & \sum_{m,n=1}^{N}\bfsfK({\bf r},{\bf R}_{m}, \omega)\cdot
\bfsfT_{mn}^{(N)}(\omega) \cdot {\bf F}^{(1)}({\bf R}_{n},\omega),
\nonumber
\end{eqnarray}
with the $N$-atom T-matrix
\begin{equation}\label{tnatomsinhomogeneous}
\bfsfT_{mn}^{(N)}(\omega) = {\hat {\bm \mu}}_{m}
T^{(N)}_{mn}(\omega){\hat {\bm \mu}}_{n} ={\hat {\bm \mu}}_{m}
T_{m}(\omega) M_{mn}^{-1}(\omega){\hat{\bm \mu}}_{n}.
\end{equation}
The $N\times N$ matrix ${\bf M}(\omega)$ is defined as
\begin{equation}\label{Mforatoms}
M_{ij}(\omega) = \left[ \delta_{ij} - (1-\delta_{ij}){\hat{\bm
\mu}}_{i}\cdot \bfsfG({\bf R}_{i},{\bf R}_{j},\omega)\cdot {\hat
{\bm \mu}}_{j} T_{j}(\omega)\right].
\end{equation}
Eqs.~(\ref{tnatomsinhomogeneous}) and (\ref{Mforatoms}) neatly sum
up infinitely many scattering events which are not described by
$\bfsfG$. Light propagation  in between the scattering off one
atom and the next one   is described by $\bfsfG$ and need not be
rectilinear, since $\bfsfG$ is the Green function of the
inhomogeneous medium.

As before, the total field ${\bf F}$ consists of the part ${\bf
E}^{(0)}$ that has not seen the atoms,  a scattered part and a
source-field part. The  field operator that describes the
scattering of light by the $N$-atom system has the form
\begin{equation}\label{LSTformatoms}
{\bf F}_{\rm scat}({\bf r},\omega)  = \sum_{m,n=1}^{N}\bfsfK({\bf
r},{\bf R}_{m},\omega)\cdot \bfsfT_{mn}^{(N)}(\omega)\cdot {\bf
E}^{(0)}({\bf R}_{n},\omega).
\end{equation}
This is a generalization of the single-atom result of
Eq.~(\ref{Fscattdef}). It describes elastic scattering as well as
resonance fluorescence off $N$ atoms in an inhomogeneous medium.
The expectation value of ${\bf F}_{\rm scat}$ depends on the
initial quantum state of light only. Similarly, for the $N$-atom
source-field that only depends on the initial atomic state, we
find
\begin{equation}\label{F2Nexact}
 {\bf F}_{\rm source}({\bf
r},\omega)
 =  \sum_{m=1}^{N}\bfsfK^{(N)}({\bf r},{\bf R}_{m},\omega)\cdot
{\bf S}_{m}(\omega),
\end{equation}
which generalizes Eq.~(\ref{sourcefielddef}). Here, $\bfsfK^{(N)}$
is a Green function of the inhomogeneous dielectric including the
$N$ atoms:
\begin{eqnarray}\label{gdielN}
&&\bfsfK^{(N)}({\bf r},{\bf r}',\omega) =  \bfsfK({\bf r},{\bf
r}',\omega)  \\ && +  \sum_{m,n=1}^{N}\bfsfK({\bf r},{\bf
R}_{m},\omega)\cdot \bfsfT^{(N)}_{mn}(\omega)\cdot \bfsfK({\bf
R}_{n},{\bf r}',\omega). \nonumber
\end{eqnarray}
For ${\bf r}$ and ${\bf r'}$ different from one of the positions
${\bf R}_{m}$, $\bfsfK^{(N)}({\bf r},{\bf r}',\omega)$ is equal to
what one would call the total Green function $\bfsfG^{(N)}$ of
both the dielectric and its guests. The solution (\ref{F2Nexact})
shows that the source-field $\bfsfK^{(N)}({\bf r},{\bf
R}_{m},\omega)\cdot {\bf S}_{m}(\omega)$ that emanates from atom
$m$ is influenced by the  positions, orientations, dipole moments
and resonance frequencies of the (N-1) other atoms. Notice that
the same $N$-atom T-matrix describes the N-atom source fields and
scattered fields. The two-atom source field will be studied in
Sec.~\ref{supertwo}.

\subsection{Interatomic interactions}\label{interatomic_inter}
In the results of Sec.~\ref{solLSSeveralatoms}, interatomic
interactions can be identified. Before doing that, we briefly
mention possible interatomic interactions that we already
neglected or that simply do not occur in our theory. In a
minimal-coupling formalism there would be a direct atom-atom
interaction in the Hamiltonian. In a multipole formalism, the only
direct interaction between neutral atoms is an interatomic
polarization energy \cite{Cohen89}. Classically, this interaction
is zero unless the smallest spheres containing the atomic charges
have nonzero overlap \cite{Cohen89}. Quantum mechanically, this
`contact energy' is negligible unless the interatomic distance is
of the order of the size of the atoms such that wave functions
overlap. We assumed that the atoms were further apart. Together
with  the fact that atoms are much smaller than the wavelength of
light, this allows us to make the dipole approximation in which
atoms are considered as point dipoles.  Therefore, direct
interactions between the atoms are absent in the dipole
Hamiltonian Eq.~(\ref{Hafquantcorrglobal}). Our approximations
make that the only interatomic interactions that we can find are
retarded dipole-dipole interactions, mediated by the
electromagnetic field.

Indeed, in Sec.~\ref{solLSSeveralatoms} interatomic interactions
showed up in the N-atom T-matrix as terms proportional to the
causal Green tensor of the medium. For two non-coinciding
positions ${\bf R}_{1}$ and ${\bf R}_{2}$, the interaction (with
dimension: [frequency]) has the form
\begin{eqnarray}\label{J12detail}
J_{12}&  = & J({\bf R}_{1},{\bf R}_{2},\omega) =
\frac{\mu_{1}\mu_{2}\omega^{2}}{\hbar\varepsilon_{0}c^{2}}\hat{{\bm
\mu}}_{1}\cdot\bfsfG({\bf R}_{1},{\bf R}_{2},\omega)\cdot\hat{{\bm
\mu}}_{2} \nonumber \\
& = & \frac{\mu_{1}\mu_{2}}{\hbar\varepsilon_{0}}
 \sum_{\lambda} \hat{\bm \mu}_{1}\cdot {\bf
f}_{\lambda}({\bf R}_{1}){\bf f}^{*}_{\lambda}({\bf
R}_{2})\cdot\hat{\bm
\mu}_{2}\frac{\omega_{\lambda}^{2}}{\omega^{2}-\omega_{\lambda}^{2}}.
\end{eqnarray}
For the latter identity, Eqs.~(\ref{Kdefgreen}) and
(\ref{Kredef2}) were used. Only after making a pole approximation
in Sec.~\ref{supertwo} will it become fully clear why we identify
precisely this expression as the dipole-dipole interaction.  Modes
with eigenfrequencies
 $\omega_{\lambda}\equiv 0$ were absent in the dipole
interaction (\ref{Hafquantcorr}) and consequently are absent in
the dipole-dipole interaction~(\ref{J12detail}).

The Green function $\bfsfG$ can be written as the sum of the
generalized transverse Green function $\bfsfG^{\rm T}$ and a
longitudinal Green function $\bfsfG^{\rm L}$ [recall
Eqs.~(\ref{Greengenlong})-(\ref{Kredef2})]. The dipole-dipole
interaction can be split into two analogous parts. The generalized
transverse part is
\begin{subequations}\label{JresJstat}
\begin{equation}\label{Jres}
J_{\rm gtrans}({\bf R}_{1},{\bf R}_{2},\omega)  =
\frac{\mu_{1}\mu_{2}\omega^{2}}{\hbar\varepsilon_{0}}
 \sum_{\lambda} \frac{\hat{\bm \mu}_{1}\cdot {\bf
f}_{\lambda}({\bf R}_{1}){\bf f}^{*}_{\lambda}({\bf
R}_{2})\cdot\hat{\bm \mu}_{2}}{\omega^{2}-\omega_{\lambda}^{2}}.
\end{equation}
It is also called the `resonant dipole-dipole interaction'  (or
RDDI).  The strongest contribution to this interaction comes from
the modes $\lambda$ with eigenfrequencies $\omega_{\lambda}$ near
$\omega$, which explains the adjective `resonant'. Notice that
$J_{\rm gtrans}$ is zero when $\omega$ is zero. The other part is
the longitudinal dipole-dipole interaction $J_{\rm long}$ that has
the mode expansion
\begin{equation}\label{Jlong}
J_{\rm long}({\bf R}_{1},{\bf R}_{2})  =
 -\frac{\mu_{1}\mu_{2}}{\hbar\varepsilon_{0}}
\sum_{\lambda}\hat{\bm \mu}_{1}\cdot{\bf f}_{\lambda}({\bf
R}_{1}){\bf f}^{*}_{\lambda}({\bf R}_{2})\cdot\hat{\bm \mu}_{2}.
\end{equation}
\end{subequations}
Notice that $J_{\rm long}$ is independent of the frequency
$\omega$. It is the generalization of the  static dipole-dipole
interaction that is well known for free space.   Both the
generalized transverse and the longitudinal dipole-dipole
interactions are given here in terms of generalized transverse
modes. Both $J_{\rm gtrans}$ and $J_{\rm long}$ are influenced by
the medium.

Both $\bfsfG^{\rm T}$ and $\bfsfG^{\rm L}$ have nonretarded dipole
terms, so that a change in a source term changes instantaneously
the longitudinal and generalized transverse fields elsewhere. It
is only their sum that is fully retarded. This is well known for
free-space Green tensors (see \cite{Craig84}) and it holds
likewise for the Green functions $\bfsfG_{\rm hom}^{\rm T}$ and
$\bfsfG_{\rm hom}^{\rm L}$ of homogeneous dielectrics as given in
the Appendix.

It might seem strange that the longitudinal interaction
Eq.~(\ref{Jlong}) is not given in terms of longitudinal modes. The
physical reason is that longitudinal modes do not couple to the
atoms in our formalism (see Sec.~\ref{Hamiltonianseveralatoms}).
Still, apart from the generalized transverse solutions ${\bf
f}_{\lambda}$ (with $\omega_{\lambda}\ne 0$) of the wave
equation~(\ref{true modes}), there are also longitudinal solutions
${\bf q}_{\nu}$ (with $\omega_{\nu} = 0$). There is a mathematical
identity that allows one to rewrite the longitudinal interaction
Eq.~(\ref{Jlong}) in terms of longitudinal modes. The identity
originates from the fact that the modes $\{{\bf f}_{\lambda}, {\bf
q}_{\nu}\}$ together span the entire  space of functions ${\bf h}$
with $\int{\mbox d}{\bf r} \varepsilon({\bf r}) |{\bf h}({\bf
r})|^{2}<\infty$. This space consists of a subspace of generalized
transverse functions and a longitudinal subspace.  The
completeness relation in the entire space reads $\sum_{\lambda}
{\bf f}_{\lambda}({\bf r}) {\bf f}^{*}_{\lambda}({\bf
r'})\varepsilon({\bf r'}) + \sum_{\nu} {\bf q}_{\nu}({\bf r}) {\bf
q}^{*}_{\nu}({\bf r'})\varepsilon({\bf r'}) = \delta({\bf r}-{\bf
r'})\bfsfI$, with $\bfsfI$ the unit tensor. It follows that for
${\bf r}\ne {\bf r'}$, one can replace $\sum_{\lambda} {\bf
f}_{\lambda}({\bf r}) {\bf f}^{*}_{\lambda}({\bf r'})$ in
Eq.~(\ref{Jlong}) by minus the sum $\sum_{\nu} {\bf q}_{\nu}({\bf
r}) {\bf q}^{*}_{\nu}({\bf r'})$. Incidentally, the longitudinal
modes ${\bf q}_{\nu}$ of the medium are different from the
free-space longitudinal modes, because of their dif\-ferent
orthogonality relations $\int{\mbox d}{\bf r} \varepsilon({\bf r})
{\bf q}_{\nu}({\bf r})\cdot{\bf q}_{\nu '}^{*}({\bf
r})=\delta_{\nu\nu'}$.

\section{Two-atom superradiance in inhomogeneous medium} \label{supertwo}
The general results of section~\ref{severalsources}  will now be
applied to two identical atoms positioned  in an inhomogeneous
dielectric. Assume that the two atoms  have identical electronic
transition frequencies $\Omega$ and dipole moments $\mu=|{\bm
\mu}|$;  their dipole orientations $\hat{{\bm \mu}}_{1}$ and
$\hat{{\bm \mu}}_{2}$ need not be identical. The source field of
this two-atom system is [see Eq.~(\ref{F2Nexact})]
\begin{equation}\label{F2N2exact}
 {\bf F}_{\rm source}({\bf
r},\omega)
 =  \bfsfK^{(2)}({\bf r},{\bf R}_{1},\omega)\cdot
{\bf S}_{1}(\omega)+ \bfsfK^{(2)}({\bf r},{\bf R}_{2},\omega)\cdot
{\bf S}_{2}(\omega).
\end{equation}
The goal is now to calculate the Green function $\bfsfK^{(2)}$ of
the dielectric including the guest atoms, in terms of the
properties of the medium and of the individual atoms.

According to Eq.~(\ref{gdielN}), the Green function $\bfsfK^{(2)}$
is known once the T matrix $T^{(2)}$ (\ref{tnatomsinhomogeneous})
is determined; $T^{(2)}$ can be found by inverting the $2\times2$
matrix $M$ (\ref{Mforatoms}), in which  the single-atom T-matrices
occur that are  given in Eq.~(\ref{tatominhomogene}) and the Green
function $\bfsfK$ of the dielectric in Eq.~(\ref{Kredef2}). It
follows that the  two-atom T-matrix is
\begin{equation}\label{T2guests}
\bfsfT^{(2)} = \frac{1/\beta}{1 -
T_{1}J_{12}^{2}T_{2}/\beta^{2}}\left(
\begin{array}{cc}
\hat{\bm \mu}_{1}\hat{\bm \mu}_{1}\beta T_{1} & \hat{\bm
\mu}_{1}\hat{\bm \mu}_{2}T_{1}J_{12}T_{2} \\
\hat{\bm \mu}_{2}\hat{\bm \mu}_{1}T_{2}J_{12}T_{1} & \hat{\bm
\mu}_{2}\hat{\bm \mu}_{2}\beta T_{2}
\end{array}
\right),
\end{equation}
with the dipole-dipole interaction $J_{12}$ defined in
Eq.~(\ref{J12detail}) and $\beta$ as
$\mu^{2}\omega^{2}/(\hbar\varepsilon_{0}c^{2})$. Each of the four
matrix elements of $\bfsfT^{(2)}$ is a dyadic of the same type as
the single-atom T-matrix (\ref{tatominhomogene}). Now abbreviate
$\bfsfK({\bf r},{\bf R}_{1},\omega)$ as $\bfsfK({\bf r}1)$ and
similarly for other terms. The Green function $\bfsfK^{(2)}({\bf
r}1)$ can be written with Eq.~(\ref{gdielN}) as
\begin{eqnarray}\label{Kgreentwopoints}
\bfsfK^{(2)}({\bf r}1) &  = &  \bfsfK({\bf r}1)\cdot
\left[\;\bfsfI + \bfsfT^{(2)}_{11}\cdot\bfsfK(11) +
\bfsfT_{12}^{(2)}\cdot\bfsfK(21)\;\right] \nonumber \\ & + &
\bfsfK({\bf r}2)\cdot\left[\;\bfsfT^{(2)}_{21}\cdot\bfsfK(11) +
\bfsfT^{(2)}_{22}\cdot\bfsfK(21)\;\right].
\end{eqnarray}
 Use
Eq.~(\ref{T2guests}) to rewrite  the T-matrix elements of
$\bfsfT^{(2)}$ in terms of the single-atom T-matrices. The first
one of the two parts of the source field (\ref{F2N2exact}) is
associated with light initially residing in atom 1. This part can
be written in terms of single-atom properties as
\begin{eqnarray}\label{K1S1rewritten}
\bfsfK^{(2)}({\bf r}1)\cdot{\bf S}_{1} & =& \left(\frac{1 +
T_{1}X_{1}/\beta}{1-T_{1}J_{12}^{2}T_{2}/\beta^{2}}\right)\times
\nonumber \\ && \left[\;\bfsfK({\bf r}1)\cdot\hat{\bm \mu}_{1}+
\bfsfK({\bf r}2)\cdot\hat{\bm
\mu}_{2}T_{2}J_{21}/\beta\;\right]S_{1},
\end{eqnarray}
with $X_{i}=X_{i}(\omega)$ as defined in
Eq.~(\ref{Xdefforoneatom}). The source field has now been
expressed in terms of the T-matrices of the individual atoms, but
it is rewarding to break up the T-matrices in parts that depend on
the medium alone and parts that depend on the atoms:
\begin{widetext}
\begin{equation}\label{K1S1rewrittenagain}
\bfsfK^{(2)}({\bf r}1)\cdot{\bf S}_{1} =
\frac{(\omega^{2}-\Omega^{2}) S_{1}\left[\;\bfsfK({\bf
r}1)\cdot\hat{\bm \mu}_{1}\left(\omega^{2}-\Omega^{2} - 2\Omega
X_{2}\right)+ \bfsfK({\bf r}2)\cdot\hat{\bm \mu}_{2}\;2\Omega
J_{12}\;\right]}{(\omega^{2}-\Omega^{2})^{2}-2\Omega\left(X_{1}+X_{2}\right)(\omega^{2}-\Omega^{2})
+ 4\Omega^{2}\left(X_{1}X_{2}-J_{12}^{2}\right)}.
\end{equation}
\end{widetext}
The denominator carries the important information about the
resonance frequencies $\Omega_{\pm}(\Omega)$ of the two-atom
system. There are two resonance frequencies near $\Omega$ and two
near $-\Omega$. If these resonance frequencies change little due
to the electromagnetic coupling with the dielectric, we can
replace the frequency dependent functions $X_{1,2}(\omega)$ and
$J_{12}(\omega)$ in the above expression by their values in
$\omega=\pm \Omega$. This is the pole approximation that we also
made for the single atom. We find the two resonance frequencies
\begin{equation}\label{resonancestwoatoms}
\Omega_{\pm}(\Omega)
 = \Omega +   \frac{X_{1} + X_{2}}{2}\pm
\sqrt{\left(\frac{X_{1}-X_{2}}{2}\right)^{2}+ J_{12}^{2}}.
\end{equation}
The other two resonance frequencies occur at
 $-\Omega_{\pm}^{*}(\Omega)$, so that all
four have negative imaginary parts.

When the atoms are far apart, then $J_{12}$ tends to zero and the
two resonance frequencies
$\Omega_{\pm}$~(\ref{resonancestwoatoms}) are simply the two
single-atom frequencies $\Omega_{1}$ and $\Omega_{2}$ with their
medium-dependent radiative shifts $\Delta_{1,2}$ and  decay rates
$\Gamma_{1,2}$. In the other extreme situation, for atoms with
parallel dipoles atoms (almost) on top of each other,
$-\mbox{Im}\Omega_{+}$ approaches twice the single-atom amplitude
decay rate, whereas $-\mbox{Im}\Omega_{-}$ has the limiting value
zero. Analogous to free space, $\Omega_{+}$ corresponds to the
superradiant state of the two-atom system in the medium, whereas
$\Omega_{-}$ is the frequency belonging to the subradiant state.

 Now rewrite Eq.~(\ref{K1S1rewrittenagain})
as a sum over individual first-order frequency poles. With
Eq.~(\ref{atomicsourceoperator}) one has
\begin{eqnarray}\label{K1S1rewrittenagain2}
\bfsfK^{(2)}({\bf r}1)\cdot{\bf S}_{1} & = &
\sum_{\pm}\left(\frac{-i\mu\omega^{2}}{4\varepsilon_{0}c^{2}\Omega_{\pm}}\right)
\nonumber \\ &&\left[\;\bfsfK({\bf r}1)\cdot\hat{\bm \mu}_{1}(1
\pm\sin\alpha)\pm \bfsfK({\bf r}2)\cdot\hat{\bm
\mu}_{2}\cos\alpha\;\right] \nonumber \\ & \times &
\left[\;(\omega+\Omega)b_{1}(0)+(\omega-\Omega)b_{1}^{\dag}(0)\right]
\nonumber \\ & \times &
\left(\frac{1}{\omega-\Omega_{\pm}}-\frac{1}{\omega+\Omega_{\pm}}\right).
\end{eqnarray}
A (complex) angle $\alpha=\alpha(\Omega)$ has been introduced
which measures the inhomogeneity of the medium as felt by the
two-atom system, in comparison with the atom-atom interaction:
\begin{equation}\label{alphainhomdef}
\sin\alpha \equiv (\Omega_{1}-\Omega_{2})/\Lambda,\quad \cos\alpha
\equiv 2J_{12}/\Lambda,
 \end{equation} with $\Lambda$ equal to
 $\sqrt{\left(\Omega_{1}-\Omega_{2}\right)^{2} + 4 J_{12}^{2}}$. When the angle
 $\alpha$ is zero (such as in free space), the atoms are said to be
 placed at equivalent positions in the medium.

 In the expression (\ref{resonancestwoatoms}) for the
resonance frequencies and in the angle $\alpha$
(\ref{alphainhomdef}), a driving term and a detuning can be
discerned. The driving term is the dipole-dipole interaction
$J_{12}$ and it signifies how important the one atom is as a light
source for the other. The term $(\Omega_{1}-\Omega_{2})/2$ is a
detuning: larger medium-induced local differences felt by the
identical atoms make the resonant transfer of a photon between
them less probable. The driving term and the detuning have the
same physical origin and can not be changed independently.  By
bringing the atoms much closer in each others near field, they
will be tuned better and interact stronger at the same time. The
outcome of the competition between medium-induced driving and
detuning will be studied in an example in
Sec.~\ref{secsuperplane}.

 The time dependence of the source field can now be calculated with an inverse Laplace
transformation. Notice that the positive-frequency poles in
Eq.~(\ref{K1S1rewrittenagain2}) have negligible residues in terms
proportional to $b_{1}^{\dag}$; similarly, negative-frequency
poles hardly contribute to terms involving the annihilation
operator $b_{1}(0)$ and can be neglected as well.  The total
source field ${\bf F}_{\rm source}({\bf r},t)$ is the field
(\ref{K1S1rewrittenagain2}) that originates from the initial
excitation of the atom labelled 1, accompanied by the source field
that originally came from the second atom:
\begin{equation}\label{F2helemaal}
{\bf F}_{\rm source}({\bf r},t) =  {\bf L}_{1}({\bf r},t) b_{1}(0)
+{\bf L}_{2}({\bf r},t) b_{2}(0) + \mbox{H.c.}
\end{equation}
 The vector ${\bf
L}_{1}$ can be written as the sum of ${\bf L}_{1+}$ and ${\bf
L}_{1-}$, with
\begin{eqnarray}\label{L1pm}
&&{\bf L}_{1\pm}({\bf r},t)  \equiv \frac{-i\mu}{4 \pi
\varepsilon_{0}c^{2}}
\int_{-\infty}^{\infty}\mbox{d}\omega\;\frac{\omega^{2}\;e^{-i\omega
t}}{\omega-\Omega_{\pm}}\cdot\frac{\omega+\Omega}{2\Omega_{\pm}} \times \\
&& \left[\;\bfsfK({\bf r},{\bf R}_{1},\omega)\cdot\hat{\bm
\mu}_{1}(1\pm\sin\alpha) \pm \bfsfK({\bf r},{\bf
R}_{2},\omega)\cdot\hat{\bm \mu}_{2}\cos\alpha\;\right]. \nonumber
\end{eqnarray}
This is the central result of this section. Vectors ${\bf
L}_{2\pm}$ can be found by interchanging the indices 1 and 2 in
the right-hand side of Eq.~(\ref{L1pm}), which also causes a sign
change in $\sin\alpha$ (\ref{alphainhomdef}).

Eq.~(\ref{L1pm}) describes the full time dependence of the source
that is excited at time $t=0$. Initially
  light has been
emitted but has not arrived at the detector yet. This initial
phase lasts a certain delay time $t_{\rm d}$, depending on the
optical path length between source and detector. The initial phase
is   followed by a transient regime, in which light that has
chosen the shortest path already arrives at the detector at ${\bf
r}$, while light that takes a longer path has not arrived yet. The
transient regime can be neglected if it lasts much shorter than
the typical atomic decay time, which is usually the case. Assuming
the same delay time for both resonance frequencies and for both
atoms, we find
\begin{eqnarray}\label{L1pm_b}
&& {\bf L}_{1\pm}({\bf r},t)   \equiv \frac{-\mu
\Omega_{\pm}(\Omega+\Omega_{\pm})
}{4\varepsilon_{0}c^{2}}\theta(t-t_{\rm d}) \; e^{-i\Omega_{\pm}
t} \times  \\ && \left[\;\bfsfK({\bf r},{\bf
R}_{1},\Omega)\cdot\hat{\bm \mu}_{1}(1\pm\sin\alpha)
    \pm \bfsfK({\bf r},{\bf
R}_{2},\Omega)\cdot\hat{\bm \mu}_{2}\cos\alpha\;\right]. \nonumber
\end{eqnarray}
In this equation we see that the source amplitudes of the atoms
are influenced by their environment. There are overall factors
which are equal for both atoms. The atoms differ in that the
source amplitude of the light that is finally emitted by the first
atom has a factor $(1\pm \sin \alpha)$ while the corresponding
factor for the second atom is $\cos\alpha$. The results for the
source field can be inserted into the intensity operator
  \cite{Loudon83}
\begin{equation}\label{intensitypoint}
I({\bf r},t) = 2 \varepsilon_{0}c {\bf E}^{(-)}({\bf
r},t)\cdot{\bf E}^{(+)}({\bf r},t),
\end{equation}
 to give the time-dependent
intensity of the light emitted by the two atoms, at a detector
position  where $\varepsilon({\bf r})$ equals unity. Suppose that
the two atoms share a single excitation so that their initial
state is the superposition
\begin{equation}\label{initialoneatomstate}
|\Psi(t=0)\rangle = \left[ p b_{1}^{\dag}(0)+
e^{i \phi}\sqrt{1-p^{2}}b_{2}^{\dag}(0)\right]|0\rangle.
\end{equation}
Then the expectation value of the intensity operator is
\begin{eqnarray}\label{intensitytwoatoms}
\langle I({\bf r},t)\rangle & = & 2\varepsilon_{0}c \biggl\{ \;
p^{2} |{\bf L}_{1}|^{2}+ (1-p^{2})|{\bf L}_{2}|^{2} \nonumber \\
& + & 2 p\sqrt{1-p^{2}}\mbox{Re}\left[\; e^{i \phi}{\bf
L}_{1}^{*}\cdot {\bf L}_{2}\;\right] \biggl\},
\end{eqnarray}
where variables $({\bf r},t)$ were dropped.

From Eqs.~(\ref{L1pm_b}) and (\ref{intensitytwoatoms}) it follows
that both atoms act as sources and superradiance can take place,
even if only the first atom is initially excited (so that $p=1$).
This point was stressed for two-atom emission in free space in
\cite{Dung00}. The time-dependent intensity that passes at ${\bf
r}$ is a complicated interference pattern of source fields emitted
by four sources: a fast(er) and a (more) slowly decaying source at
${\bf R}_{1}$, and also a fast and a slow source at the atomic
position ${\bf R}_{2}$. The photon is shared by and exchanged
between the atoms until it is finally emitted, via either the fast
superradiant or the slow subradiant decay process. Amplitudes of
the fast and slow sources originating from an initially unexcited
atom depend both on the interaction between the atoms and on their
medium-induced detuning.

\section{Application: Superradiance near a partially reflecting plane}\label{secsuperplane}
Our multiple-scattering formalism will now be applied to the
situation of two identical atomic dipoles in the vicinity of an
infinitely thin plane that partially reflects light. We are
interested in medium-induced spontaneous-emission rates, Lamb
shifts, interatomic interactions, and sub- and superradiant decay
rates. These quantities of interest can be expressed in terms of
the Green tensor of the  medium. In a recent paper \cite{Wubs04},
we already developed a method to efficiently calculate the Green
tensor of a medium consisting of one `plane scatterer' and of
several parallel planes, but we only used it to calculate
spontaneous-emission rates.   For a more detailed discussion of
the plane-scatterer model and the method to calculate the Green
tensor, we refer to \cite{Wubs04}. Below, we give a brief outline
of our calculations. We will then focus on those aspects of our
results that we believe are generic for many more inhomogeneous
dielectrics.

 The atomic dipoles are assumed identical and parallel to each other (${\bm \mu}_{1}={\bm \mu}_{2}$),
 and parallel  to the plane.
Moreover, assume the atomic positions ${\bf
R}_{i}=(x_{i},y_{i},z_{i})$ to be confined to the line
$x_{i}=y_{i}=0$. The plane is assumed perpendicular to the
$z$-axis. The interatomic interaction Eq.~(\ref{J12detail}) can
then be written as
 \begin{equation}\label{Jplane}
 J_{12}=J({\bf R}_{1}, {\bf R}_{2}, \Omega) = \Gamma_{0}\left(\frac{3
 c}{4\Omega}\right)\int_{0}^{\infty}\mbox{d}k_{\parallel}
 k_{\parallel}\left( G^{ss} + G^{vv} \right).
 \end{equation}
 Here, $G^{ss}$ stands for the component of the Green tensor $G(k_{\parallel},z_{1}, z_{2},\Omega)$
 that describes propagation of $s$-polarized light, while $G^{vv}$ describes $p$-polarized light \cite{Wubs04}.
 For a single plane, we have
\begin{eqnarray}\label{Gss}
 && G^{ss}(k_{\parallel},z_{1}, z_{2},\Omega)  =   G_{0}^{ss}(k_{\parallel},z_{1}, z_{2},\Omega)  \\
 &&  +
 G_{0}^{ss}(k_{\parallel},z_{1}, z_{\rm
 plane},\Omega)T^{ss}(k_{\parallel},\Omega)G_{0}^{ss}(k_{\parallel},z_{\rm plane},
 z_{2},\Omega). \nonumber
 \end{eqnarray}
The free-space tensor component $G_{0}^{ss}(k_{\parallel},z_{1},
z_{2},\Omega)$ equals $\exp(i k_{z}|z_{1}-z_{2}|)/(2 i k_{z})$,
with the wave vector $k_{z}$ defined as
$\sqrt{(\Omega/c)^{2}-k_{\parallel}^{2}}$. The T-matrix
$T^{ss}(k_{\parallel},\Omega)$ of the plane for $s$-polarized
light has the form $-[(D_{\rm
eff}(\Omega/c)^2)^{-1}-i/(2k_z)]^{-1}$. The plane is fully
characterized by the single parameter $D_{\rm eff}$, which we call
its `effective thickness'. We choose the value $D_{\rm eff}=0.23
\lambda$. With this choice,  $32\%$ of $s$-polarized light is
transmitted through the plane when averaged over $4\pi$ incoming
angles. Higher values of $D_{\rm eff}$ give less transmission. For
the Green tensor component $G^{vv}$  in Eq.~(\ref{Jplane}) one can
write an expression analogous to Eq.~(\ref{Gss}): in the
right-hand side of Eq.~(\ref{Gss}), the components $G_{0}^{ss}$
must be replaced by $G_{0}^{vv}=(k_{z}c/\Omega)^{2}G_{0}^{ss}$,
and $T^{ss}(k_{\parallel},\Omega)$ by
$T^{vv}(k_{\parallel},\Omega)=-[(D_{\rm eff}(\Omega/c)^2)^{-1}-i
k_{z}c^{2}/(2\Omega^{2})]^{-1}$.

If in the integral Eq.~(\ref{Jplane}) the in-plane wave vector
$k_{\parallel}$ becomes larger than $\Omega/c$, then the wave
vector $k_{z}$ becomes purely imaginary and equal to $i\kappa$,
with $\kappa$ equal to $\sqrt{k_{\parallel}^{2}-(\Omega/c)^{2}}$.
The semi-infinite integration interval in Eq.~(\ref{Jplane})
therefore falls apart into two parts: a radiative part with
$k_{\parallel}$ between $0$ and $\Omega/c$, and an evanescent part
with $k_{\parallel}$ from $\Omega/c$ onwards. The evanescent part
of the integral is purely real, except that there is a purely
imaginary contribution from a pole in $T^{ss}$ at $\kappa=D_{\rm
eff}(\Omega/c)^{2}/2$. This pole corresponds to an $s$-polarized
guided mode.  Near the pole, the real evanescent part of the
integral over the s-wave integrand in Eq.~(\ref{Jplane}) must be
taken as a Cauchy principal-value integral. There is no
corresponding pole in $T^{vv}$. The evanescent part of the
integral for $p$-polarized light is purely real and can be
evaluated numerically right away. All relative errors in our
numerical results are smaller than $10^{-6}$.

\begin{figure}[t]
 \begin{center}
 {\includegraphics[width=80mm,height=64mm]{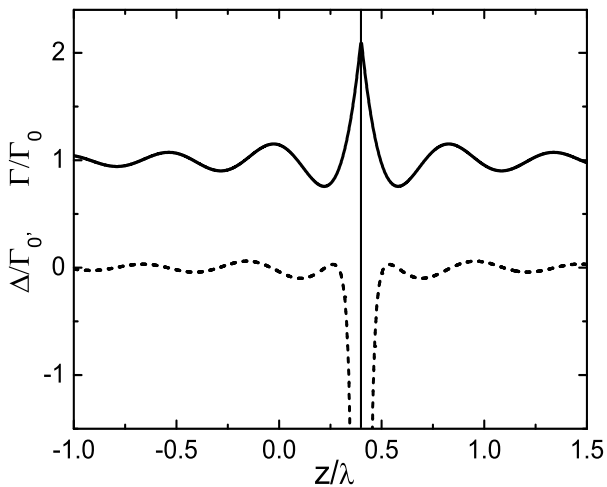}}
 \end{center}
 \caption{Spontaneous-emission rate $\Gamma$ (solid line) and Lamb shift $\Delta$ (dashed line) as a function of the position of an atom near
  a partially reflecting plane. The effective thickness of the plane is $D_{\rm eff}=0.23\lambda$. The atomic dipole moment points parallel to the
  plane.  The plane is positioned  at $z_{\rm plane}=0.4 \lambda$, to make comparisons with later figures easier.
  Both $\Gamma$ and $\Delta$ are given in units of $\Gamma_{0}$,
  and the positions $z$ are scaled to the wavelength $\lambda=2\pi c/\Omega$ of the emitted
  light. The period of the damped oscillations in both $\Gamma$ and $\Delta$ is  $\lambda/2$.}\label{fig1}
 \end{figure}

The single-atom spontaneous-decay rate
Eq.~(\ref{decaysinglesource})   can be found from the interaction
Eq.~(\ref{J12detail}) through the relation $\Gamma({\bf
R}_{1},\Omega)= -2\mbox{Im}J_{12}({\bf R}_{1},{\bf
R}_{1},\Omega)$, while the the Lamb shift
Eq.~(\ref{reallevelshift}) follows from $\Delta({\bf
R}_{1},\Omega)=\mbox{Re}[J_{12}({\bf R}_{1},{\bf
R}_{1},\Omega)-J_{12}^{(0)}({\bf R}_{1},{\bf R}_{1},\Omega)]$.
Fig.~\ref{fig1} shows how single-atom properties are modified by
the presence of the plane.
The figure shows a peak in the decay rate near the plane due to
emission into the guided modes \cite{Wubs04}. Away from the plane,
the decay rate shows damped oscillations towards the free-space
decay rate $\Gamma_{0}$. There are two oscillations per wavelength
$\lambda$, a characteristic also well-known for spontaneous
emission near a perfect mirror \cite{Milonni94}. The Lamb shift
shows similar damped oscillations around $\Delta=0$ away from the
plane. At distances less than $\lambda/10$ the shift becomes
strongly negative and it actually diverges to minus infinity. The
atom is attracted to the plane \cite{Milonni94}, but here we
assume atomic positions to be fixed.

\begin{figure}[t]
 \begin{center}
 {\includegraphics[width=80mm,height=64mm]{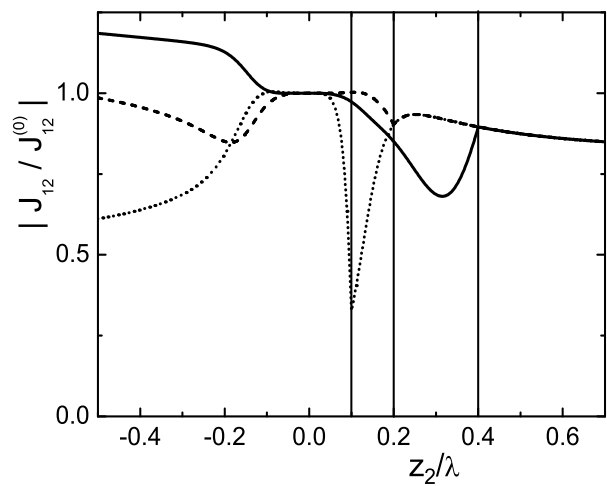}}
 \end{center}
 \caption{Absolute values of interatomic interactions $J_{12}({\bf R}_{1},{\bf R}_{2},\Omega)$ near
 a partially reflecting plane,
 scaled to the free-space interaction strength
 $|J_{12}^{(0)}({\bf R}_{1},{\bf R}_{2},\Omega)|$. The plane is as
 in Fig.~\ref{fig1}.
 The first atom's position is fixed in the origin.
 The second atom travels along the line $(x_{2}=y_{2}=0,z_{2})$. The atomic dipoles
 point in the same directions, parallel to the plane. The three graphs differ in the position of the plane with respect
 to the first atom. Solid line: $z_{\rm
 plane}/\lambda=0.4$; dashed line: $z_{\rm
 plane}/\lambda=0.2$; dotted line: $z_{\rm
 plane}/\lambda=0.1$. (All three planes are shown, but in each
 case considered only a single plane is present.)
  }\label{fig2}.
 \end{figure}

In Fig.~\ref{fig2}
we present dipole-dipole interactions for two atoms near a plane.
The first atom is kept fixed in the origin, the distance of the
plane to this first atom is chosen, and then the absolute value of
the interatomic interaction $J_{12}$ is plotted as a function of
the position of the second atom, relative to the free-space value
$|J_{12}^{(0)}|$. The interaction is the sum of  radiative and
evanescent interactions,  of  both  $s$-polarized and
$p$-polarized light. The figure shows that for $z_{2}$ approaching
$z_{1}=0$, the
 relative difference between $|J_{12}|$ and the (divergent) free-space interaction
 strength $|J_{12}^{(0)}|$ becomes negligible,
 irrespective of the  position of the plane. This holds independently of the reflectivity
 of the plane (not shown in Fig.~\ref{fig1}). Interestingly,
 the dipole-dipole interaction  is also
 independent of the position of the plane (but not of its reflectivity) if the plane
 stands
 in between the two atoms.  In other words, with the atomic positions fixed at either side of the plane,
 one can move the plane back and forth without changing the interatomic interaction.
 This fact can be read off from
 Fig.~\ref{fig2} for $z_{2}/\lambda>0.4$, where the three graphs (corresponding to
 three plane positions) overlap. It can also be understood from
 the form of the interaction in Eq.~(\ref{Jplane}), because all
 terms in the interaction either depend on $|z_{2}-z_{1}|$ or on
 $(|z_{1}-z_{\rm plane}|+|z_{2}-z_{\rm plane}|)$. For $z_{2}<0$ and $|z_{2}|\gg
 \lambda$, the relative interaction $|J_{12}/J_{12}^{(0)}|$ approaches a constant value, which can be
 either larger or smaller than unity, depending on the distance of the plane to atom~1.
 As a check on our calculations (not shown), we found that interatomic
 interactions vanished (as expected) when an almost ideal mirror (a plane with
 $D_{\rm eff}=100\lambda$) is placed in between them.

\begin{figure}[t]
 \begin{center}
 {\includegraphics[width=80mm,height=64mm]{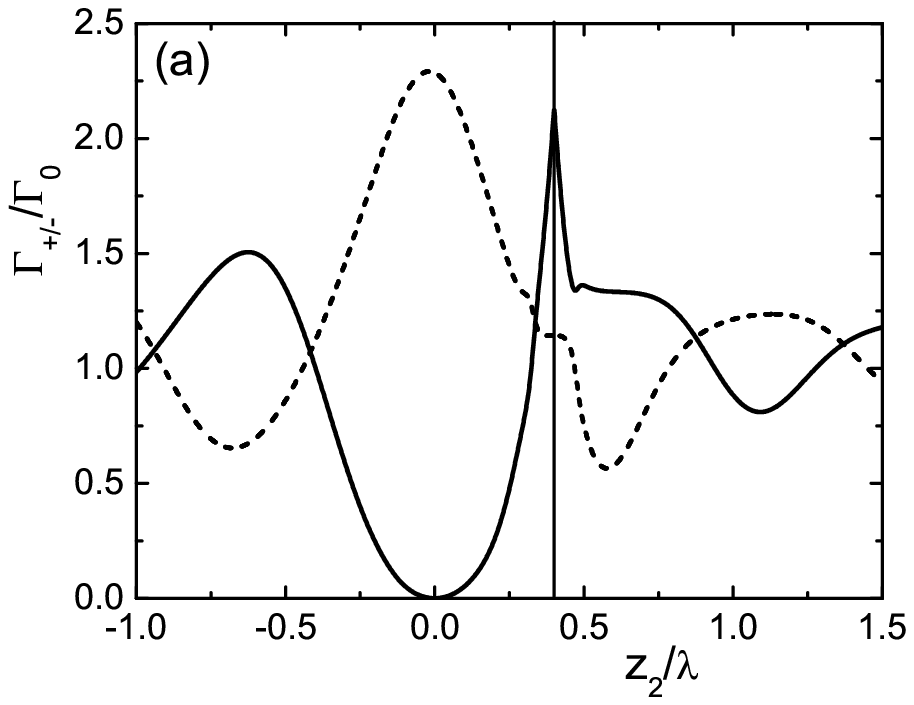}}
 {\includegraphics[width=80mm,height=64mm]{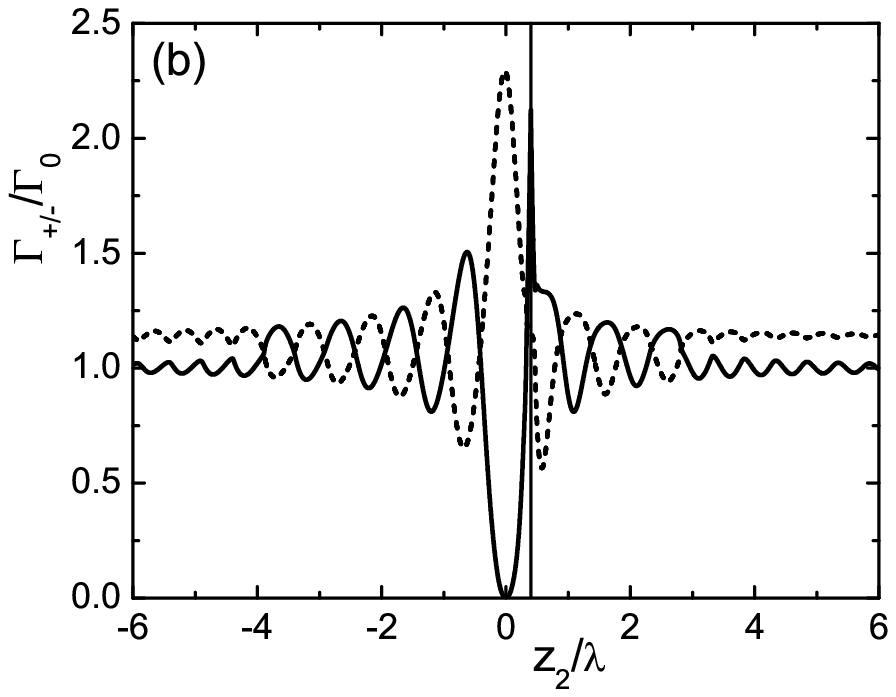}}
 \end{center}
 \caption{Subradiant and superradiant decay rates for two atoms near a partially reflecting plane,
 as a function of the position of the second atom. The situation is as in Fig.~\ref{fig2} with the plane
 fixed at $z_{\rm plane}/\lambda=0.4$. Fig.~(a) zooms in around the plane, showing a vanishing subradiant decay rate
 $\Gamma_{-}$ (solid line)  as the second atom approaches the first one in the origin. The superradiant decay
 rate $\Gamma_{+}$ (dashed line)
 becomes more than twice the single-atom decay rate $\Gamma_{0}$.
 Both $\Gamma_{-}$ and $\Gamma_{+}$
 show perturbed
 oscillations on the scale of $\lambda$. Fig.~(b) zooms out to larger distances, showing a cross-over regime
 between damped oscillations with a period $\lambda$ and more distant oscillations with a period $\lambda/2$.}\label{fig3}
 \end{figure}

Fig.~\ref{fig3}(a) shows two-atom  superradiant and subradiant
decay rates, as modified by the presence of the plane. The plots
are based on Eqs.~(\ref{resonancestwoatoms}) and (\ref{Jplane}).
The complex square root in Eq.~(\ref{resonancestwoatoms}) has
solutions that differ by an overall minus sign. Care was taken to
choose the solution from the same branch as we varied the position
of the second atom.
Without the plane, one would have had
$\Gamma_{0,\pm}=\Gamma_{0}\mp 2\mbox{Im} J_{12}^{(0)}$. With the
plane, the medium-induced detuning becomes negligible as $z_{2}$
approaches $z_{1}$. Then $\Gamma_{-}/\Gamma_{0}$ vanishes, as in
free space. The corresponding small-distance limit of
$\Gamma_{+}/\Gamma_{0}$ is not equal to 2 as for free space, but
rather twice the single-atom decay rate $\Gamma=1.14\Gamma_{0}$ in
the presence of the plane. If the second atom moves towards the
mirror, then the medium-induced detuning (see Fig.~\ref{fig1})
grows fast while the dipole-dipole interaction (Fig.~\ref{fig2})
decreases. With Eq.~(\ref{resonancestwoatoms}) we then find that
$\Gamma_{-}\simeq \Gamma_{2}$ and $\Gamma_{+}\simeq \Gamma_{1}$.
Indeed, for $z_{2}$ closer than $\lambda/10$ to the plane,
$\Gamma_{-}$ follows the single-plane emission rate of
Fig.~\ref{fig1}, while $\Gamma_{+}$ equals $\Gamma({\bf
R}_{1},\Omega)=1.14\Gamma_{0}$. With atom 2 so close to the plane,
superradiance is completely absent, even though the identical
atoms are less than half a wavelength apart. For
$z_{2}>0.5\lambda$ or $z_{2}<0.3\lambda$, detuning has become less
important and the decay rates follow (not quite sinusoidal) damped
oscillations. Their period is $\lambda$, as it is for
superradiance in free space.

 Fig.~\ref{fig3}(b) again shows super- and subradiant decay rates,
 now also for larger interatomic distances.  For
 $-4\lesssim z_{2}/\lambda \lesssim 3$, the rates $\Gamma_{\pm}$ exhibit the same damped
 oscillations with  period $\lambda$ that we also saw in
 fig.~\ref{fig3}(a). However, for $z_{2}/\lambda$ smaller than -4 or larger than 3,
$\Gamma_{\pm}$ show {\em two} oscillations per wavelength, like we
saw for the single-atom decay rate in Fig.~\ref{fig1}. Hence we
can identify a rather sharp  cross-over regime at a few
wavelenghts away from the plane between superradiance and
single-atom emission. For larger distances, again medium-induced
detuning dominates the dipole-dipole interaction. Indeed for large
distances we see the same behavior as very close to the plane,
namely that $\Gamma_{-}$ approaches $\Gamma_{2}$ (which at these
positions almost equals $\Gamma_{0}$) while $\Gamma_{+}$ has the
limiting value $\Gamma_{1}=1.14\Gamma_{0}$. In the cross-over
regime,  $|J_{12}|$ has the same order of magnitude as the
detuning $|\Delta_{1}-\Delta_{2}-i(\Gamma_{1}-\Gamma_{2})/2|$.

 If one puts the
plane closer to the first atom, then this atom becomes further
detuned from its free-space properties. The cross-over should then
take place with the second atom at shorter distances where the
interaction is still stronger. This we have verified (not shown).
At a fixed frequency, the spatial intervals in which superradiance
occurs therefore depend on three distances, namely the interatomic
distance and the distances between each atom and the plane.

\begin{figure}[t]
 \begin{center}
 {\includegraphics[width=80mm,height=64mm]{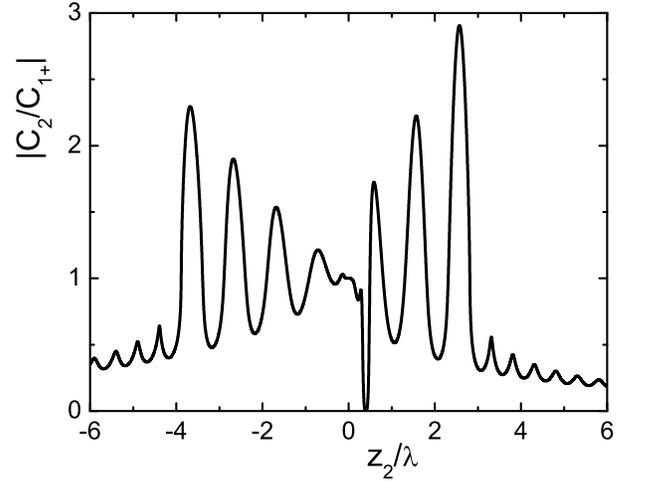}}
 \end{center}
 \caption{Absolute value of the source amplitude $C_{2}=\cos\alpha$ of the second atom,
 divided by the source amplitude $C_{1+}=(1+\sin\alpha)$ of the
 first atom. The situation is as in Fig.~\ref{fig3}. The amplitudes are associated with superradiant emission, see
 Eq.~(\ref{L1pm_b}).
  }\label{fig4}
 \end{figure}

Not only the  super- and the subradiant emission rates are
influenced by the presence of the plane, but also the source
amplitudes of the two atoms are modified, shown in
Eq.~(\ref{L1pm_b}): if initially only the first atom is excited,
then the source amplitude of the second atom is modified by a
factor $C_{1\pm}\equiv 1\pm \sin\alpha$, for superradiant ($+$)
and subradiant ($-$) decay, respectively. Atom 2 gets a factor
$C_{2} \equiv \cos \alpha$ for both  decay processes.
 Fig.~\ref{fig4} shows  $|C_{2}/C_{1+}|$ as $z_{2}$ is varied.
The emission rates of Fig.~\ref{fig3} and the relative amplitudes
in Fig.~\ref{fig4} show the same cross-over regions between
oscillations with periods $\lambda/2$ and  $\lambda$. When the two
atoms coincide, the detuning vanishes and $|C_{2}/C_{1+}|$ equals
unity. (For free space, $|C_{2}/C_{1+}|$ equals unity everywhere,
even if the atoms do not coincide.) The coinciding atoms are
equivalent and superradiance can occur. On the other hand, close
to the plane at $|z_{2}/\lambda-0.4|\lesssim 0.1$, atom 2 is
strongly detuned and $|C_{2}/C_{1+}|$ vanishes: the second atom
emits none of the light initially residing in the first one and
superradiance does not occur.

Fig.~\ref{fig4} also shows that at larger distances
($z_{2}\lesssim -4$ or $z_{2}\gtrsim 3$),
 detuning is again strong enough to make
emission by the second atom less probable than in free space. At
these larger distances, the medium-induced detuning suppresses the
net transfer of light from atom 1 to atom 2 and superradiance does
not occur.  In the two intervals
 $-4<z_{2}\lambda<0.3$ and $0.5<z_{2}/\lambda<3$ where
 superradiance does occur, we see that the peaks of the relative
 source amplitude $|C_{2}/C_{1+}|$ are higher than unity. There the probability
 that light initially residing on the first atom is finally emitted by the second one is higher than in free space.
 The peaks of Fig.~(\ref{fig4}) correspond to positions of the second atom for which
  most light is finally emitted by the second atom,
 although initially only the first atom was excited. Interestingly, the peaks of $|C_{2}/C_{1+}|$  become higher as the
 second atom moves away from the first. The highest peaks occur when the complex-valued dipole-dipole interaction
 (almost) exactly
 compensates
 the complex-valued  detuning. (Such a resonant situation does not exist for identical atoms in free
 space.) For larger $|z_{2}|$, the dipole-dipole interaction
 becomes too weak to compensate for the detuning and
 Fig.~\ref{fig4} shows  very abrupt transitions from superradiance
 to single-atom emission on both sides of the plane.

\section{Conclusions, discussion, and outlook}\label{quantcorrconc}
In this paper, a multiple-scattering theory was set up with at its
heart the Lippmann-Schwinger equation~(\ref{eqforfieldexactlipp})
that describes the electromagnetic field operators in an
inhomogeneous dielectric with guest atoms present. We solved the
LS equation exactly in terms of the properties of the atoms and
the Green tensors of the medium, both when one and when several
guest atoms are present. The solution for the electric field
operator has three parts: a part that has not seen the guest
atoms, a part that describes the scattering by the resonant atoms,
and a part that describes the atoms as sources.

Our formalism is a generalization of an already existing
point-scattering formalism for classical waves. The generalization
is twofold: first, our formalism is valid not only for free space
but for atoms in all dielectrics that can be described
macroscopically in terms of a real relative dielectric constant
$\varepsilon({\bf r})$. Second, it is a multiple-scattering theory
in quantum optics rather than classical optics.  In relation to
this point we find the double nature of atoms both as scatterers
and as sources of light. The formalism is quantum mechanical in
the sense that it can describe the propagation and scattering of
nonclassical sources of light. These  can be either external or
atomic sources. In quantum optics, the medium must be described
with more care, just like the quantum and classical descriptions
of a beam splitter differ \cite{Mandel95}. As for the beam
splitter, classical light sources give classical measured fields
in our formalism, since we described the guest atoms as harmonic
oscillators.

A nice feature of the LS equation~(\ref{eqforfieldexactlipp}) is
that it follows exactly from a dipole Hamiltonian that is the
result of a canonical quantization theory. The Hamiltonian
describes guest atoms microscopically and treats the dielectric
macroscopically. The atomic dipoles do not couple to the electric
field operator ${\bf E}$ but rather to a field operator that we
call ${\bf F}$ and that includes the atom's own polarization
field. For free space this is a well known result. We find that
the propagator for the field ${\bf F}$ in our LS equation is not
the ordinary Green tensor $\bfsfG$, but rather a Green tensor that
we called $\bfsfK$. There exists a simple relation~(\ref{Kredef2})
between $\bfsfG$ and $\bfsfK$ for an arbitrary dielectric.
$\bfsfG$ can be split into the generalized transverse Green tensor
$\bfsfG^{\rm T}$ that propagates the vector potential ${\bf A}$,
and the longitudinal Green tensor $\bfsfG^{\rm L}$.

In the Appendix we showed that  the volume-integrated electric
field~(\ref{Eintegrated}) produced in free space by an atomic
dipole is equal to minus one third of its polarization field. This
is an operator relation at finite frequency. A different
(incorrect) relation would have resulted if  the field ${\bf F}$
had been interpreted as the electric-field operator. We have not
come across other work that addresses the relation between  the
dipole interaction, the occurrence of $\bfsfK$ rather than
$\bfsfG$ in a multiple-scattering theory, and the
volume-integrated electric field around a dipole. In this respect,
our formalism   also sheds new light on quantum optics in free
space.

The infinitely sharp single-atom resonance in the potential
$\bfsfV$  obtains a radiative shift and a width in the T-matrix
$\bfsfT$. In our formalism, the position-dependent shift and decay
rate are the summed effects of infinitely many light-scattering
events off the atomic potential.  The scattered-field operator for
a single atom contains two parts: an elastic-scattering term and a
term describing resonance fluorescence. Direct interatomic
interactions are absent in the dipole Hamiltonian
(\ref{Hafquantcorr}). Dipole-dipole interactions appear
`dynamically' in the solutions of the Lippmann-Schwinger equation
for several atoms. An inhomogeneous medium modifies both the
longitudinal and the generalized transverse dipole-dipole
interactions, see Eq.~(\ref{JresJstat}).

The multiple-scattering formalism has been used to study
superradiance in an  inhomogeneous medium. The often dominant
electronic component to inhomogeneous broadening was neglected in
order to focus on photonic effects. As an application, we studied
how dipole-dipole interactions and two-atom superradiance are
influenced by a partially reflecting plane. We found
position-dependent modifications of dipole-dipole interactions.
For our choice of parameters,  the plane suppresses superradiance
if one of the atoms is very close or very far from the plane. Both
atoms will then emit as if alone. For intermediate distances,
two-atom sub- and superradiance will occur. Due to the plane,
emission rates are modified and so are the relative amplitudes of
the atomic sources. Interestingly, we found that medium-induced
complex detuning
 can lead to enhanced transfer of light
from the one atom to the other, before superradiant emission
occurs. Also, we found  sharp cross-overs between spatial
intervals where superradiance occurs (with decay rates oscillating
once per wavelength) and single-atom emission (two oscillations
per wavelength).

The length of the intervals in which superradiance occurs depends
on the atomic positions with respect to each other and to the
plane. This length could be called a ``perpendicular coherence
length''. This would complement the concept of a transverse
coherence length (or effective mode radius)
\cite{DeMartini90,Ujihara02}. The latter concept is used in the
analysis of cooperative emission in a planar microcavity when the
atoms have the same $z$-coordinate, but have different coordinates
in another direction. An important difference between the two
lengths is that only the perpendicular coherence length is
influenced by medium-induced detuning.

We believe that our results for cooperative emission near the
plane are generic and that similar cross-over regions will occur
in more complex dielectrics.  Still, it would be interesting to
study
 the influence of other dielectric structures on multi-atom processes,
Bragg mirrors for example, or `optical corrals'
\cite{Colas01,Chicanne02}.  Photonic crystals are also very
interesting media, for which superradiance has only been studied
in an isotropic model \cite{Kurizki88,John95,Vats98} where all
position dependence is neglected. Like near a plane, superradiance
inside a real photonic crystal will be influenced by
medium-induced detuning.   As another application of our
formalism, statistical distributions of optical proximity
resonances of many-atom systems can be studied, to find analogies
and differences  in inhomogeneous optical and electronic systems
\cite{Rusek00}.

We made a pole approximation  in a late stage of our formalism,
after which we found exponential atomic decay. The pole
approximation no longer holds when the atom-field interaction
becomes strong \cite{Kurizki96}. The approximation also breaks
down if local densities of states jump steeply as a function of
frequency near the atomic transition frequency. There is a current
debate whether pole approximations will break down at the band
edges of realistic three-dimensional photonic crystals
\cite{Li00}, like it is found for the isotropic model
\cite{John90}. In principle, our formalism could also be used
without making the pole approximation.

Our theory is valid if frequency dispersion of the medium can be
neglected. Now single-atom emission rates only depend on one
frequency of the medium, so that dispersion is not important. On
the other hand, radiative line shifts, interatomic interactions,
and hence superradiant decay rates do depend on all frequencies of
the medium. In our example of two atoms near a plane, the
immediate vicinity of the atoms was free space. However, for atoms
embedded in a $\varepsilon \ne 1$ part of a medium,  line shifts
would diverge unless frequency dispersion of the medium is taken
into account \cite{Milonni99}. This will also be the case for
position-dependent radiative shifts in photonic crystals
\cite{Vats02}. It will be interesting to study the influence of
frequency dispersion of the medium on cooperative atomic emission,
for example based on Refs.~\cite{Savasta02,Dung02}.

\section*{Acknowledgements}
We would like to thank Peter Lodahl, Allard Mosk, Gerard Nienhuis,
Rudolf Sprik, Bart van Tiggelen,  and Willem Vos for stimulating
discussions. This work is part of the research program of the
Stichting voor Fundamenteel Onderzoek der Materie, which is
financially supported by the Nederlandse Organisatie voor
Wetenschappelijk Onderzoek.

\appendix
\section*{Appendix: Homogeneous dielectric}

\subsection*{1. Delta and Green functions}\label{Homocase}
For a homogeneous dielectric with refractive index $n$, the
property `generalized transverse' reduces to transverse in the
ordinary sense. The medium is translational  invariant, so that
${\bm \delta}^{\rm T,L}({\bf r}, {\bf r'})={\bm \delta}^{\rm
T,L}({\bf r}- {\bf r'})$. The transverse and longitudinal delta
functions appearing in Eqs.~(\ref{deltagegtrans}) and
(\ref{deltaLdef}) now become
\begin{subequations}\label{deltashom}
\begin{eqnarray}\label{deltatranslongfree}
{\bm \delta}_{\rm hom}^{\rm T}({\bf r}) & = &
\frac{2}{3}\delta({\bf r})\bfsfI - \frac{1}{4 \pi r^{3}}(\bfsfI-3
\hat{\bm r}\otimes \hat{\bm r}) \label{deltatransfree}\\ {\bm
\delta}_{\rm hom}^{\rm L}({\bf r}) & = & \frac{1}{3}\delta({\bf
r})\bfsfI + \frac{1}{4 \pi  r^{3}}(\bfsfI-3 \hat{\bm r}\otimes
\hat{\bm r}),\label{deltalongfree}
\end{eqnarray}
\end{subequations}
 where $\hat{\bm r}$ is defined as
${\bf r}/|{\bf r}|$, the unit vector in the direction of ${\bf
r}$. The sum of the transverse and the longitudinal delta function
is simply $\delta({\bf r})\bfsfI$, since their `dipole' parts
cancel. Notice that $n$ does not enter these delta functions. The
derivation follows the free-space treatment \cite{Craig84}.

The dyadic Green function $\bfsfG_{\rm hom}({\bf r}, {\bf
r'})=\bfsfG_{\rm hom}({\bf r}- {\bf r'})$ for the homogeneous
medium is the sum of a transverse and a longitudinal part. The
transverse part is \cite{DeVries98a}
\begin{subequations}\label{GTL}
\begin{eqnarray}\label{g0transpoint}
\bfsfG_{\rm hom}^{\rm T}({\bf r},\omega) & = & - \frac{\bfsfI - 3
\hat{\bf r}\otimes\hat{\bf r}}{4 \pi (n\omega/c)^{2} r^{3}}
 \\ &  - & \frac{e^{i n \omega r/c}}{4 \pi r} \left[P(i n
\omega r/c)\bfsfI + Q(i n \omega r/c)\hat{\bf r}\otimes\hat{\bf
r}\right],\nonumber
\end{eqnarray}
with the function $P(z)$ defined as $(1-z^{-1} + z^{-2})$ and
$Q(z)$ as $(-1+3z^{-1}-3z^{-2})$. With the use of the
definition~(\ref{GLdef}) of the longitudinal Green function and
the transverse delta function (\ref{deltatransfree}), the
longitudinal Green function is found to be
\begin{equation}\label{ghomlongpoint}
\bfsfG_{\rm hom}^{\rm L}({\bf r},\omega)  =   \frac{\bfsfI - 3
\hat{\bf r}\otimes\hat{\bf r}}{4 \pi (n \omega/c)^{2} r^{3}} +
\frac{\delta({\bf r})}{3 (n \omega/c)^{2}}\bfsfI.
\end{equation}
\end{subequations}
The delta-function term in $\bfsfG_{\rm hom}^{\rm L}$ appears
naturally and there was no need to add it `by hand' as is done
elsewhere \cite{Jackson75,DeVries98a}.

\subsection*{2. Volume-integrated dipole field}\label{Volumeintegralsection}
The rigorous multiple-scattering formalism of
Sec.~\ref{scatsesingle} with the Green functions $\bfsfK$ will now
be used to calculate the volume integral of the electric-field
operator ${\bf E}$ in terms of the atomic polarization fields
$\sum_{m}{\bf P}_{m}$ of Eq.~(\ref{pgpoint}), with the volume
taken over a small sphere enclosing an atom.

With the help of the Eqs. (\ref{dotbfourier}),
(\ref{dotbdagfourier}), and the definitions of the source fields
(\ref{atomicsourceoperator}) and potentials (\ref{Vmnorwa}), the
polarization field in frequency space can be related to other
operators as
\begin{equation}\label{pginfourterms}
{\bf P}_{m}(\omega) =
-\left(\frac{\varepsilon_{0}c^{2}}{\omega^{2}}\right)\left[\;{\bf
S}_{m}(\omega) + \bfsfV_{m}(\omega)\cdot{\bf F}({\bf R}_{m},
\omega)\;\right].
\end{equation}
There exists therefore a simple relationship between the field
{\bf F} and the polarization fields [use
Eq.~(\ref{eqforfieldexactlipp})]
\begin{equation}\label{fandprelated}
{\bf F}({\bf r},\omega) = {\bf E}^{(0)}({\bf r},\omega) -
\frac{\omega^{2}}{\varepsilon_{0} c^{2}}\sum_{m=1}^{N} \bfsfK({\bf
r},{\bf R}_{m},\omega)\cdot {\bf P}_{m}(\omega).
\end{equation}
This equation is still valid for all inhomogeneous dielectrics.
Now assume that the sources are in free space. Consider the
volume-integral of the field ${\bf F}$ over a small sphere
(denoted by $\odot$) containing only the source at ${\bf R}_{m}$,
at its center.  The integral is determined by the free-space Green
function $\bfsfK_{0}({\bf r},{\bf R}_{m},\omega)$ for positions
${\bf r}$ close to ${\bf R}_{m}$ (see
Eqs.~(\ref{Kredefsecondline}) and (\ref{Kredef2})). The transverse
Green function $\bfsfG_{0}^{\rm T}({\bf r}-{\bf R}_{m},\omega)$ in
$\bfsfK_{0}$ has a vanishing contribution to the integral, since
its pole goes as $|{\bf r}-{\bf R}_{m}|$ at short distances (see
Eq.(\ref{g0transpoint}). The dipole part of the transverse delta
function (\ref{deltatransfree}) has a vanishing angle-integral
over the sphere and does not contribute either. What remains is
the delta-function part of the transverse delta function, which
gives the radius-independent result
\begin{equation}\label{fintegrated}
\int_{\odot}\mbox{d}{\bf r}\;{\bf F}({\bf r},\omega) = \frac{2}{3
\varepsilon_{0}}{\bf P}_{m}(\omega).
\end{equation}
Now the subtlety becomes important that the field ${\bf F}= -{\bf
D}({\bf R}) /[\varepsilon_{0}\varepsilon({\bf R})]$ is equal to
the electric field ${\bf E}$ everywhere except at the positions of
the guest atoms [see Eq.~(\ref{Fdef})]. The expression in Eq.
(\ref{fintegrated}) is therefore not equal to the
volume-integrated electric field. With the definitions of the
 fields ${\bf D}$ and  ${\bf F}$ given in Sec.~\ref{Hamiltonianseveralatoms},
 one obtains the
relation
\begin{equation}\label{Eintegrated}
\int_{\odot}\mbox{d}{\bf r}\;{\bf E}({\bf r},\omega) = -\frac{1}{3
\varepsilon_{0}}{\bf P}_{m}(\omega).
\end{equation}
The static and classical version of this `sum rule' is presented
for example in \cite{Jackson75}. There, and more recently in
\cite{DeVries98a}, a delta function is added by hand to the static
dipole field or to the longitudinal Green function. In contrast,
Eq.~(\ref{Eintegrated}) was found here as an operator relation
without adding any terms by hand.

The interpretation of the field to which a dipole couples is not
just a matter of choice in the present formalism. If one wrongly
identifies ${\bf F}$ as the electric field ${\bf E}$ but correctly
derives the relation (\ref{Kredef}) or (\ref{Kredef2}) between
$\bfsfK$ and $\bfsfG$, then the wrong volume-integrated electric
field $2{\bf P}_{m}(\omega)/(3\varepsilon_{0})$ would have
resulted. The delta function term that is the difference between
$\bfsfG$ and $\bfsfK$ in (\ref{Kredef2}) and the difference
between the field operators ${\bf E}$ and ${\bf F}$ have the same
physical origin: the atomic polarization field.

Still, there is nothing truly quantum mechanical about the sum
rule (\ref{Eintegrated}). In a classical canonical theory, one
would find the same  dipole coupling $-\bfmu\cdot{\bf F}$ and
Green function $\bfsfK$. However, a canonical formalism is usually
by-passed in classical optics. It is then assumed that a classical
dipole couples to the classical electric field and furthermore
that light propagates from a source according to the Green
function $\bfsfG$ rather than $\bfsfK$. By summing
Eqs.~(\ref{g0transpoint}) and (\ref{ghomlongpoint}) for $n=1$, one
finds that $\bfsfG_{0}$ na\-tu\-ral\-ly has the correct
delta-function term to produce Eq.~(\ref{Eintegrated}). Therefore,
although following a less rigorous procedure, one has the luck
that there is no need to add terms by hand in order to derive
Eq.~(\ref{Eintegrated}) classically.

\end{document}